\documentclass[journal]{IEEEtran}
\pdfoutput=1

\usepackage{graphicx}
\usepackage{cuted}
\usepackage[cmex10]{amsmath}
\usepackage{cases}
\usepackage{amsmath}
\usepackage[tight,footnotesize]{subfigure}
\usepackage{amsthm}                 
\usepackage{cite}
\usepackage{amssymb}
\usepackage{threeparttable}
\usepackage{utfsym}
\usepackage{bbding}
\usepackage{svg}
\usepackage{booktabs}
\usepackage{bbm, dsfont}
\usepackage{mathrsfs}
\usepackage[scr=rsfs,cal=boondox]{mathalfa}

\usepackage{yfonts}
\usepackage{algorithm}
\usepackage{algorithmic}
\usepackage{multirow}
\usepackage{xcolor}
\usepackage{CJK}
\usepackage{subeqnarray}
\usepackage{cases}
\usepackage{longtable}
\usepackage{enumitem}
\usepackage{bm}
\usepackage{tikz}
\usepackage{comment}

\ifCLASSINFOpdf
\else
\fi

\begin{document}
%



%
\title{A unified framework for STAR-RIS coefficients optimization}
\author{Hancheng~Zhu,~\IEEEmembership{Student~Member,~IEEE,}
	Yuanwei~Liu,~\IEEEmembership{Senior~Member,~IEEE,}
	Yik-Chung~Wu,~\IEEEmembership{Senior~Member,~IEEE,}
	Vincent~K.~N.~Lau,~\IEEEmembership{Fellow,~IEEE}
	\thanks{Hancheng Zhu and Yik-Chung Wu are with the Department of Electrical and Electronic Engineering, The University of Hong Kong, Hong Kong (email: u3006551@hku.hk, ycwu@eee.hku.hk).}
	\thanks{Yuanwei Liu are with the School of Electronic Engineering
		and Computer Science, Queen Mary University of London, London E1 4NS,
		U.K (email: yuanwei.liu@qmul.ac.uk).}
	\thanks{Vincent~K.~N.~Lau is with Hong Kong University of Science and Technology, Hong Kong, China (email:eeknlau@ee.ust.hk).}}
\maketitle
\begin{abstract}
	Simultaneously transmitting and reflecting (STAR) reconfigurable intelligent surface (RIS), which serves users located on both sides of the surface, has recently emerged as a promising enhancement to the traditional reflective only RIS. Due to the lack of a unified comparison of communication systems equipped with different modes of STAR-RIS and the performance degradation caused by the constraints involving discrete selection, this paper proposes a unified optimization framework for handling the STAR-RIS operating mode and discrete phase constraints. With a judiciously introduced penalty term, this framework transforms the original problem into two iterative subproblems, with one containing the selection-type constraints, and the other subproblem handling other wireless resource. Convergent point of the whole algorithm is found to be at least a stationary point under mild conditions. As an illustrative example, the proposed framework is applied to a sum-rate maximization problem in the downlink transmission. Simulation results show that the algorithms from the proposed framework outperform other existing algorithms tailored for different STAR-RIS scenarios. Furthermore, it is found that 4 or even 2 discrete phases STAR-RIS could achieve almost the same sum-rate performance as the continuous phase setting, showing for the first time that discrete phase is not necessarily a cause of significant performance degradation.
\end{abstract}
\begin{IEEEkeywords}
simultaneously transmitting and reflecting reconfigurable intelligent surface (STAR-RIS), operating mode constraint, discrete phase constraint, unified framework.
\end{IEEEkeywords}
\section{introduction}
The development of meta-surface/tunnel diode technology provides a feasible avenue for the realization of reflective intelligent surfaces (RIS) in forthcoming communication systems. RIS, characterized by its cost-effectiveness and remarkable scalability, leverages an array of reflective meta-surfaces affixed to walls or surfaces. These meta-surfaces are endowed with the capability to manipulate phase shifts and amplitudes, thereby enhancing the propagation of signals \cite{Wu:19,Kudathanthirige:20,Basar:19,Chen:19}. By redirecting incoming signals towards the receiver, the RIS establishes a virtual direct pathway connecting the transmitter and receiver, effectively circumventing physical obstructions \cite{Renzo:19}. This innovation significantly facilitates wireless transmissions across a multitude of scenarios \cite{Li:22,Yang:22}.

Traditionally, RIS solely functions to reflect signals, limiting its coverage to receivers positioned on the same side as the transmitters. A pioneering approach to achieve full ${\rm{360}}^\circ$ coverage is the concept of Simultaneously Transmitting and Reflecting (STAR) RIS, which has recently emerged \cite{Xu:21,Wu:21,Niu:21}. This innovative paradigm enables the concurrent reflection and transmission (refraction) of incident signals, effectively catering to users situated on both sides of the surface.

To fully unlock the potential in STAR-RIS technology, the transmitting and reflecting coefficients need to be properly optimized. Compared to conventional RISs, the optimization of the STAR-RIS coefficients is subjected to a set of intricate constraints. For example, there are three distinct operational modes - energy splitting (ES), mode switching (MS) and time switching (TS). In ES mode, the principles of energy conservation and lossless power considerations dictates that the sum power of transmitting and reflecting coefficients confines to unity. On the other hand, operation under MS mode allocates each constituent element of the STAR-RIS to either transmission or reflection, resulting in a mixed-integer programming (MIP) problem, which is NP-hard even under the simple quadratic objective function~\cite{Alom:17}. To address this intricacy, a prevalent strategy involves the conversion of integer constraints into a penalty term and adds it to the objective function \cite{Perera:22,Mu:22}. As the penalty weight increases, each STAR-RIS element is forced to either transmission or reflection. In TS mode, all elements of the STAR-RIS are dedicated to fully transmission or reflection at a time. Therefore, the amplitude of the STAR-RIS transmission and reflection coefficients are constrained to be one. In this mode, the optimization problem is to determine the phase of the STAR-RIS coefficients and the optimal time allocation for both reflection and transmission \cite{Mu:22}.

Furthermore, recent investigations have unveiled a distinctive phase coupling phenomenon in ES STAR-RIS. This intriguing property dictates that the difference between the phases of reflection and transmission is consistently maintained at ${\pi \mathord{\left/ {\vphantom {\pi {\rm{2}}}} \right. \kern-\nulldelimiterspace} {\rm{2}}}$~\cite{Liu:22a,Zhong:22,Liu:22b}. This constraint does not exists in reflective only RIS so its exploration is still at its infancy. The earliest attempt to incorporate this constraint within STAR-RIS optimization employs the element-wise alternating optimization (AO) method~\cite{Liu:22b}. In this approach, optimization is carried out for individual RIS elements sequentially so that each subproblem has only two potential choices (either transmission phase surpasses the reflection phase by ${\pi  \mathord{\left/
		{\vphantom {\pi  {\rm{2}}}} \right.
		\kern-\nulldelimiterspace} {\rm{2}}}$ or vice versa), and the solution with better objective value is then chosen as the subsequent iteration point. Although this leads to a duplication in the number of optimization subproblems, it circumvents the combinatorial challenge of jointly determining the coupled-phase options of all STAR-RIS elements. More recently,~\cite{Wang:23} proposes to transform the coupled-phase constraint and the previously mentioned sum power constraint into penalty terms. Then, AO was employed to alternatively optimize the amplitude and phase of the STAR-RIS coefficients. Compared with elementwise-AO method, the solution quality of this penalty-based algorithm is guaranteed when the problem satisfies the Mangasarian-Fromovitz constraint qualification (MFCQ) conditions.
	
	While the prevailing model in existing STAR-RIS research adopts continuous phase shifts, it is essential to acknowledge that practical limitations from finite control signal resolution or hardware constraints leads to a finite number of permissible phases~\cite{Chen:21}. Such discrete phase phenomenon has emerged as one of the basic models in reflective only RIS. It is reasonable to anticipate that STAR-RIS encounters the same discrete phase constraints. However, the combinatorial nature of discrete phase optimization has an exponential compleixty order with the number of STAR-RIS elements. This inherent complexity renders a direct approach to addressing this constraint computationally unfeasible for STAR-RIS with massive elements. Therefore, numerous extant studies would simply ignore this discrete phase constraint during optimization, and then apply quantization to the resultant continuous phase~\cite{Gao:21}. Regrettably, such a quantization-based strategy lacks a performance guarantee for solution quality. Moreover, this strategy leads to a significant performance loss if the number of allowable phases is small (e.g., 2 or 4). 

Through the preceding deliberations, it is evident that different constraints in STAR-RIS were handled with a multitude of optimization techniques, which makes the comparison of different types of STAR-RIS difficult. Furthermore, when multiple mentioned constraints exist concurrently, it is unclear which of the existing methods can be generalized to such scenario. To fill this gap, this paper for the first time establishes an innovative unified framework for handling the operating mode and discrete phase constraints. Through the introduction of auxiliary variables for the STAR-RIS coefficients, we strategically transform the original problem into two distinct subproblems. One subproblem is dedicated to constraints involving discrete selection, while the other addresses the sum power constraint along with the additional wireless resource constraints. This strategic decoupling enables the derivation of a closed-form global optimal solution for the subproblem associated with selection constraints, which facilitates the proof of the solution quality of the proposed framework. Specifically, the converged solution is guaranteed to be at least a stationary point under mild conditions. To the best of our knowledge, this is an inaugural work that simultaneously provides solution quality guarantee in various STAR-RIS configurations, even under discrete phases. 

To illustrate the efficiency of the proposed framework, we apply it to a downlink STAR-RIS assisted sum-rate maximization system. Through this framework, the solutions for various types of STAR-RIS can be obtained simultaneously. Simulation results show that the proposed framework outperforms other existing methods. Furthermore, the proposed framework could have performance akin to that of continuous phase setting even as sparse as four or two discrete phases, which overthrows the conventional notion that discrete phase is to be blamed for performance degradation. Since the proposed framework separates the handling of selection-type constraints arising from STAR-RIS and the optimization of other wireless resource, it can be easily extended to communication scenarios involving other objective functions and radio resources. Such versatility substantially mitigates the anticipated challenges associated with future resource allocation problem under STAR-RIS.

The rest of the paper is organized as follows. The general STAR-RIS aided communication model, its penalty framework, and the conditions for solution quality guarantee are presented in Section II. Then the closed-form solution of the selection related subproblem is derived in Section III. The downlink STAR-RIS assisted sum-rate maximization is formulated and solved according to the proposed penalty framework in Section IV. Simulation results are provided in Section V and conclusions are drawn in Section VI.

\begin{figure} 
	\centering
	\includegraphics[width=1\linewidth]{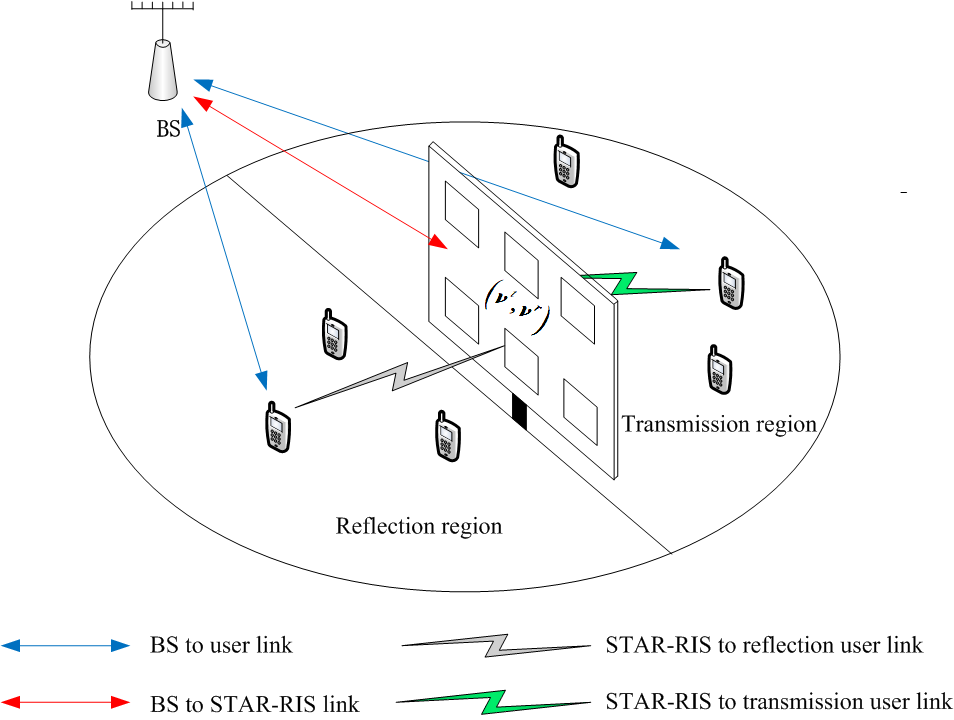}
	\caption{A typical STAR-RIS assisted communication system}
	\label{fig:graph}
\end{figure}

\section{A general STAR-RIS optimization model}
\newtheorem{myLem}{\textbf{Lemma}}
\newtheorem{myPro}{\textbf{Proposition}}
\newtheorem{myCol}{\textbf{Corollary}}
\newtheorem{myExa}{\textbf{Example}}
\newtheorem{myRem}{\textbf{Remark}}
\newtheorem{myAss}{\textbf{Assumption}}
\newtheorem{myDef}{\textbf{Definition}}
\newtheorem{mySta}{\textbf{Statement}}
\newtheorem{myProP}{\textbf{Property}}
\newcommand*{\circled}[1]{\lower.7ex\hbox{\tikz\draw (0pt, 0pt)%
		circle (.5em) node {\makebox[1em][c]{\small #1}};}}

We consider a communication system assisted by a STAR-RIS as shown in Fig. 1. Users are distributed on both sides of STAR-RIS with $M$ elements. We denote the transmission coefficient and the reflection coefficient at the ${m^{th}}$ STAR-RIS element as $v_m^{\mathcal t} \in {\rm{{\mathbb C}}}$ and $v_m^{\mathcal r} \in {\rm{{\mathbb C}}}$ respectively. Due to the existence of various types of STAR-RIS, the STAR-RIS coefficients are subjected to different constraints, which are discussed below.

\subsection{Modeling of STAR-RIS Constraints}

\newcommand{\tabincell}[2]{
	\begin{tabular}{@{}#1@{}}#2\end{tabular}
}
\begin{table*}
	\centering
	\begin{threeparttable}[b]
		\caption{Typical STAR-RIS models}
		\label{tab:test2}
		\begin{tabular}{|c|c|c|c|c|c|c|}
			\hline
			& & & & & &\\[-7pt]
			\tabincell{c}{STAR-RIS\\Case Index}&\tabincell{c}{Operating\\ mode}&\tabincell{c}{Amplitude\\constraint}&\tabincell{c}{Coupled-phase constraint\\$\angle v_m^{\mathcal t} - \angle v_m^{\mathcal r} =$\\$\left\{ {{\pi  \mathord{\left/
							{\vphantom {\pi  2}} \right.
							\kern-\nulldelimiterspace} 2}\left( {\bmod 2\pi } \right),{{ - \pi } \mathord{\left/
							{\vphantom {{ - \pi } 2}} \right.
							\kern-\nulldelimiterspace} 2}\left( {\bmod 2\pi } \right)} \right\}$}&\tabincell{c}{Discrete phase constraint\\$\left\{ {0,{{2\pi } \mathord{\left/
							{\vphantom {{2\pi } L}} \right.
							\kern-\nulldelimiterspace} L}, \cdots {{2\pi \left( {L - 1} \right)} \mathord{\left/
							{\vphantom {{2\pi \left( {L - 1} \right)} L}} \right.
							\kern-\nulldelimiterspace} L}} \right\}$}&\tabincell{c}{Time allocation\\ constraint\\${\lambda ^{\mathcal t}} + {\lambda ^{\mathcal r}} = 1$}&\tabincell{c}{Existing works\\ employing\\this model}\\
			\hline
			& & & & & &\\[-7.5pt]
			1&\multirow{2}{*}{TS}&\multirow{2}{*}{1}&\usym{2718}&\usym{2718}&\CheckmarkBold&~\cite{Mu:22,Katwe:23,Qin:23}\\ \cline{1-1} \cline{4-7}
			& & & & & &\\[-6pt]
			2& &
			&\usym{2718}&\CheckmarkBold&\CheckmarkBold&~\cite{Du:23,Wuyu:22}\\
			\hline
			& & & & & &\\[-6pt]
			3&\multirow{2}{*}{MS}&\multirow{2}{*}{$\left\{ {0,1} \right\}$}&\usym{2718}&\usym{2718}&\usym{2718}&~\cite{Perera:22,Ni:22,Qin:23}\\ \cline{1-1} \cline{4-7}
			& & & & & &\\[-6pt]
			4& & &\usym{2718}&\CheckmarkBold&\usym{2718}&\tabincell{c}{No existing\\ work considered\\ this model yet}\\
			\hline
			& & & & & &\\[-6pt]
			5& \multirow{4}{*}{ES}&\multirow{4}{*}{$\left[ {0,1} \right]$}&\usym{2718}&\usym{2718}&\usym{2718}&~\cite{Zhang:22b,Fang:23,Zhang:22}\\ \cline {1-1} \cline{4-7}
			& & & & & &\\[-6pt]
			6& & &\usym{2718}&\CheckmarkBold&\usym{2718}&~\cite{Abrar:23,Zhao:22}\\ \cline {1-1} \cline{4-7}
			& & & & & &\\[-6pt]
			7& & &\CheckmarkBold&\usym{2718}&\usym{2718}&\tabincell{c}{~\cite{Katwe:23,Zhai:23,Wang:23}\\~\cite{Liu:22b,Zhang:23}}\\ \cline {1-1} \cline{4-7}
			& & & & & &\\[-6pt]
			8& & &\CheckmarkBold&\CheckmarkBold&\usym{2718}&~\cite{Zhang:23}\\ 
			\hline
		\end{tabular}
	\end{threeparttable}
\end{table*}

There are three types of constraints for STAR-RISs, namely operating mode constraint, lossless power constraint and phase constraint. These constraints are detailed as follows. 

\begin{figure} 
	\centering
	\includegraphics[width=1\linewidth]{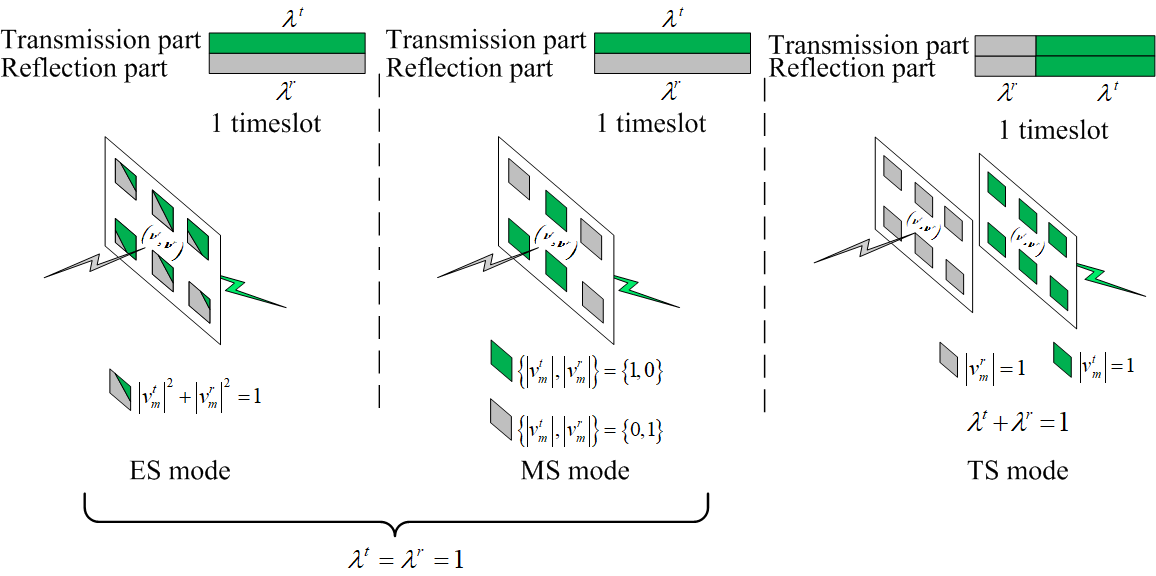}
	\caption{Three operating modes of the STAR-RIS}
	\label{fig:graph}
\end{figure}
\begin{enumerate}[leftmargin=0.5cm]
\item \textbf{Operating mode constraint}. There are commonly three operating modes of STAR-RIS~\cite{Mu:22}. The first one is ES mode, where the energy of incident signal is split between reflection and transmission, giving rise to the constraint $\left| {v_m^{\mathcal t}} \right|$, $\left| {v_m^{\mathcal r}} \right| \in \left[ {0,1} \right]$. The second one is MS mode where each STAR-RIS element is dedicated to reflection or transmission, giving rise to the constraint $\left| {v_m^{\mathcal t}} \right|$, $\left| {v_m^{\mathcal r}} \right| \in \left\{ {0,1} \right\}$. The third one is TS mode which divides the transmission interval into two regions, with one region dedicated to reflection while the other to transmission. Let ${\lambda^{\mathcal t}} \ge 0$ and ${\lambda^{\mathcal r}} \ge 0$ denoting the percentage of the time allocated to transmission period and reflection period respectively, we have ${\lambda^{\mathcal t}} + {\lambda^{\mathcal r}} = 1$. On the other hand, for ES and MS modes, as both transmission and reflection occupy the whole interval, we can set ${\lambda^{\mathcal t}} = {\lambda^{\mathcal r}} = 1$.
\item \textbf{Lossless power constraint}. It is usually assumed that the metasurface is lossless. Hence, in ES and MS mode, the reflected energy plus transmitted energy must be equal to the incident signal energy. This gives rise to the constraint ${\left| {v_m^{\mathcal t}} \right|^2} + {\left| {v_m^{\mathcal r}} \right|^2} = 1$. On the other hand, in TS mode, since transmission or reflection is the only operation in a certain time interval, lossless constraint means ${\left| {v_m^{\mathcal t}} \right|}={\left| {v_m^{\mathcal r}} \right|} = 1$. Fig. 2 illustrates the STAR-RIS operating in different modes and the corresponding constraints.
\item \textbf{Phase constraint}. It is known that the STAR-RIS phase may not take infinite resolution in practice. In this case, $\angle v_m^{\mathcal t}$, $\angle v_m^{\mathcal r} \in \left\{ {0,{{2\pi } \mathord{\left/
			{\vphantom {{2\pi } L}} \right.
			\kern-\nulldelimiterspace} L}, \cdots {{2\pi \left( {L - 1} \right)} \mathord{\left/
			{\vphantom {{2\pi \left( {L - 1} \right)} L}} \right.
			\kern-\nulldelimiterspace} L}} \right\}$, where $L$ is the number of allowable phases. More recently, a new coupled-phase model is proposed in~\cite{Liu:22a,Zhong:22,Liu:22b}, which states that a necessary condition for a proper STAR-RIS in ES mode is $\left| {v_m^{\mathcal t}} \right|\left| {v_m^{\mathcal r}} \right|\cos \left( {\angle v_m^{\mathcal t} - \angle v_m^{\mathcal r}} \right) = 0$. This is equivalent to $\angle v_m^{\mathcal t} - \angle v_m^{\mathcal r} \in \left\{ {{\pi  \mathord{\left/
			{\vphantom {\pi  2}} \right.
			\kern-\nulldelimiterspace} 2}\left( {\bmod 2\pi } \right),{{ - \pi } \mathord{\left/
			{\vphantom {{ - \pi } 2}} \right.
			\kern-\nulldelimiterspace} 2}\left( {\bmod 2\pi } \right)} \right\}$ if $\left| {v_m^{\mathcal t}} \right|$ and $\left| {v_m^{\mathcal r}} \right|$ are both non-zero.
\end{enumerate}

Different types of STAR-RIS are the results of mix and match of above constraints, and they are summarized in Table I. Notice that MS with coupled-phase is just the basic MS STAR-RIS since MS requires $\left| {v_m^{\mathcal t}} \right|$, $\left| {v_m^{\mathcal r}} \right| \in \left\{ {0,1} \right\}$. Together with the lossless power constraint ${\left| {v_m^{\mathcal t}} \right|^2} + {\left| {v_m^{\mathcal r}} \right|^2} = 1$, we must have either $\left| {v_m^{\mathcal t}} \right| = 0$ or $\left| {v_m^{\mathcal r}} \right| = 0$. This makes the coupled-phase constraint $\left| {v_m^{\mathcal t}} \right|\left| {v_m^{\mathcal r}} \right|\cos \left( {\angle v_m^{\mathcal t} - \angle v_m^{\mathcal r}} \right) = 0$ automatically satisfied. For the TS mode, since only reflection phase or transmission phase of each STAR-RIS element is used in a certain time period, coupled-phase constraint could not exist. Therefore, MS and TS STAR-RIS with coupled-phase (with or without discrete phase constraint) are not feasible and we do not list them in Table I.

Let $\bm z$ denote other communication resources to be optimized and ${{\bm v}^{\mathcal t}} = {\left[ {v_1^{\mathcal t}, \cdots ,v_M^{\mathcal t}} \right]^T}$, ${{\bm v}^{\mathcal r}} = {\left[ {v_1^{\mathcal r}, \cdots ,v_M^{\mathcal r}} \right]^T}$ be the shorthand notations for the collection of $\left\{ {v_m^{\mathcal t}} \right\}_{m = 1}^M$ and $\left\{ {v_m^{\mathcal r}} \right\}_{m = 1}^M$ respectively, a general optimization problem involving STAR-RIS can be formulated as
\begin{subequations}\label{eq:4}
	\begin{align}
		&\mathop {\min }\limits_{\left\{ {{\bm z},{{\bm v} ^{{\mathcal t}}},{{\bm v} ^{{\mathcal r}}},{{\lambda} ^{{\mathcal t}}},{{\lambda} ^{{\mathcal r}}}} \right\}}\;\; {\rm{{\cal F}}}\left( {{\bm z},{{\bm v} ^{{\mathcal t}}},{{\bm v} ^{{\mathcal r}}},{{\lambda} ^{{\mathcal t}}},{{\lambda} ^{{\mathcal r}}}} \right)\label{eq:4a}\\
		&\;\;\;\;\;\;\;\;s.t.\;\;\;{\lambda^{\mathcal t}}{\left| {v_m^{\mathcal t}} \right|^2} + {\lambda^{\mathcal r}}{\left| {v_m^{\mathcal r}} \right|^2} = 1,\label{eq:4b}\\
		&\;\;\;\;\;\;\;\;\;\;\;\;\;\;\left\{ \begin{array}{l}\vspace{1ex}
			\left| {v_m^{\mathcal t}} \right| = \left| {v_m^{\mathcal r}} \right| =1,\left\{ {{{\lambda} ^{{\mathcal t}}},{{\lambda} ^{{\mathcal r}}}} \right\}\ge 0,\quad\;\;\;\;\;{\rm if\;TS}\\\vspace{1ex}
			\left| {v_m^{\mathcal t}} \right|,\left| {v_m^{\mathcal r}} \right| \in \left\{ {0,1} \right\},{{\lambda} ^{{\mathcal t}}}={{\lambda} ^{{\mathcal r}}}=1, \;\;\;\;\;{\rm if\;MS}\\ 
			\begin{array}{l}
				\angle v_m^{\mathcal t} - \angle v_m^{\mathcal r} \in \left\{ {{\pi  \mathord{\left/
							{\vphantom {\pi  2}} \right.
							\kern-\nulldelimiterspace} 2}\left( {\bmod 2\pi } \right),} \right.\\
				\left. { - {\pi  \mathord{\left/
							{\vphantom {\pi  2}} \right.
							\kern-\nulldelimiterspace} 2}\left( {\bmod 2\pi } \right)} \right\},{\lambda ^{\mathcal t}} = {\lambda ^{\mathcal r}} = 1,
			\end{array}\;\;\;\;\;{\rm if\;ES}
		\end{array} \right.\label{eq:4c}\\
		&\;\;\;\;\;\;\;\;\;\;\;\;\;\;\;\angle v_m^{\mathcal t},\angle v_m^{\mathcal r} \in \left\{ {0,{{2\pi } \mathord{\left/
					{\vphantom {{2\pi } L}} \right.
					\kern-\nulldelimiterspace} L}, \cdots ,{{2\pi \left( {L - 1} \right)} \mathord{\left/
					{\vphantom {{2\pi \left( {L - 1} \right)} L}} \right.
					\kern-\nulldelimiterspace} L}} \right\},\label{eq:4d}\\
		&\;\;\;\;\;\;\;\;\;\;\;\;\;\;\left( {\bm z,{{\bm v}^{\mathcal t}},{{\bm v}^{\mathcal r}},{{\lambda}^{\mathcal t}},{{\lambda}^{\mathcal r}}} \right) \in {\bm \Omega}.\label{eq:4e}
	\end{align}
\end{subequations}
where ${{\rm{{\cal F}}}}$ is the objective function and is assumed to be bounded from below, which is a trivial assumption since the problem in \eqref{eq:4} is a minimization problem. The objective function in \eqref{eq:4} can represent different forms of system performance. For example, it can be power consumption~\cite{Mu:22,Liu:22b}, or mean square error function~\cite{Wuyu:22}. On the other hand, the sum-rate~\cite{Perera:22,Wang:23}, spectral efficiency~\cite{Fang:23} and secrecy capacity~\cite{Zhang:23,Zhang:22} are also frequently used optimization objective, but they need a negative sign if they are used in \eqref{eq:4} since these criteria should be maximized instead of minimized. ${\bm \Omega}$ is the coupled constraint set of $\bm z$, $\bm v^{\mathcal t}$, $\bm v^{\mathcal r}$, $\lambda^{\mathcal t}$, $\lambda^{\mathcal r}$. Constraint \eqref{eq:4b} is a general expression covering all three modes of STAR-RIS. In particular, when $\left| {v_m^{\mathcal t}} \right| = \left| {v_m^{\mathcal r}} \right| =1$, \eqref{eq:4b} reduces to time allocation constraint ${\lambda^{\mathcal t}} + {\lambda^{\mathcal r}} = 1$ in the TS mode. On the other hand, in ES or MS mode, they do not involve the allocated time variables ${\lambda^{\mathcal t}}$, ${\lambda^{\mathcal r}}$ and therefore setting ${\lambda^{\mathcal t}}={\lambda^{\mathcal r}}=1$ make \eqref{eq:4b} reduces to the lossless constraint ${\left| {v_m^{\mathcal t}} \right|^2} + {\left| {v_m^{\mathcal r}} \right|^2} = 1$. Notice that the ES mode constraint $\left| {v_m^{\mathcal t}} \right|$, $\left| {v_m^{\mathcal r}} \right| \in \left[ {0,1} \right]$ is implicitly included in \eqref{eq:4b}, therefore it is not listed in \eqref{eq:4}. The discrete phase constraint \eqref{eq:4d} are compatible with the coupled-phase constraint (third line of \eqref{eq:4c}) if $L$ is an even number greater than 2. For example, when $L=4$, $\angle \varphi _m^{\mathcal t} \in \left\{ {0,{\pi  \mathord{\left/
			{\vphantom {\pi  2}} \right.
			\kern-\nulldelimiterspace} 2},\pi ,{{3\pi } \mathord{\left/
			{\vphantom {{3\pi } 2}} \right.
			\kern-\nulldelimiterspace} 2}} \right\}$ and the coupled constraint would make the reflection phase $\angle \varphi _m^{\mathcal r} = \angle \varphi _m^{\mathcal t}  \pm {\pi  \mathord{\left/
		{\vphantom {\pi  2}} \right.
		\kern-\nulldelimiterspace} 2}$, which is still in $\left\{ {0,{\pi  \mathord{\left/
			{\vphantom {\pi  2}} \right.
			\kern-\nulldelimiterspace} 2},\pi ,{{3\pi } \mathord{\left/
			{\vphantom {{3\pi } 2}} \right.
			\kern-\nulldelimiterspace} 2}} \right\}$. Similar observations can be made as long as $L>2$ and is an even number. This condition can be easily satisfied when the number of information bit for phase control is larger than $1$.

Existing works only solve special cases of \eqref{eq:4}. For example, when only \eqref{eq:4b} and the second constraint in \eqref{eq:4c} are included, this corresponds to the typical MS model, and a penalty term ${\left| {v_m^{\mathcal t}} \right|^2} - \left| {v_m^{\mathcal t}} \right|$ is commonly introduced to relax the $\left\{ {0,1} \right\}$ constraint into $\left[ {0,1} \right]$~\cite{Perera:22,Ni:22,Qin:23}. On the other hand, when only \eqref{eq:4b} with ${\lambda^{\mathcal t}}={\lambda^{\mathcal r}}=1$ and \eqref{eq:4d} are included, \eqref{eq:4} becomes the discrete ES STAR-RIS, and a common approach is to relax the discrete phase temporarily and then quantizing the continuous-valued result into discrete phase~\cite{Abrar:23}. Furthermore, with \eqref{eq:4b} and the first case of \eqref{eq:4c}, the general formulation \eqref{eq:4} becomes the TS STAR-RIS optimization problem. Since the transmission and reflection coefficients are not coupled in TS STAR-RIS and the time allocation constraint is convex, techniques such as semidefinite relaxation (SDR)~\cite{Mu:22} and gradient descent (GD) method~\cite{Du:23} for traditional reflection only RIS can be adopted to handle the phase optimization. Recently, a coupled-phase STAR-RIS model was considered in~\cite{Liu:22b,Wang:23}, where \eqref{eq:4b} and the third line of \eqref{eq:4c} are included. To overcome the difficulty introduced by the elementwise coupled-phase constraint,~\cite{Liu:22b} proposes an elementwise-AO method, and~\cite{Wang:23} puts forward a penalty-based algorithm by moving both \eqref{eq:4b} and the second line of \eqref{eq:4c} into a penalty term. Also,~\cite{Zhang:23} uses the same penalty form to solve the secrecy beamforming problem under the coupled-phase STAR-RIS.

Although existing works tackles different special cases of problem \eqref{eq:4} by invoking different optimization schemes, there is a lack of unified framework for solving the general problem including diverse types of STAR-RIS in \eqref{eq:4}. Especially, the discrete phase-shift constraint \eqref{eq:4d} is dominantly handled by quantization, which may introduce noticeable performance loss when the number of allowable phases is small (e.g., $L=2$ or $4$), not to mention the lack of solution quality guarantee by such approach. Coming up with a unified framework not only saves the effort of finding optimization algorithms for various special cases falling into the form of \eqref{eq:4}, but also facilitates the comparison among various STAR-RIS models in a particular communication scenario. 
\subsection{A Penalty-based Reformulation of \eqref{eq:4}}
Notice that problem \eqref{eq:4} is challenging to solve for STAR-RIS coefficients ${{\bm v}^{\mathcal t}}$ and ${{\bm v}^{\mathcal r}}$ since the constraint \eqref{eq:4c} contains binary selection. More specifically, the second constraint and the third constraint in \eqref{eq:4c} are the binary selection for the amplitude and the difference of the two phases, respectively. Moreover, possible occurance of the discrete phases \eqref{eq:4d} also makes \eqref{eq:4} a mixed integer optimization problem.

To tackle this problem, we propose to employ auxiliary vectors ${{\bm \varphi} ^{\mathcal t}}$, ${{\bm \varphi} ^{\mathcal r}}$$ \in {{\rm{{\mathbb C}}}^{M \times 1}}$ together with a penalty term to handle constraints \eqref{eq:4c} and \eqref{eq:4d}. The reformulated problem is written as
\begin{subequations}\label{eq:5}
	\begin{align}
	&\mathop {\min }\limits_{\left\{ {\scriptstyle {\bm z},{{\bm v}^{{\mathcal t}}},{{\bm v}^{{\mathcal r}}},{\bm \varphi} ^{{\mathcal t}},\hfill\atop
			\scriptstyle{{\bm \varphi} ^{{\mathcal r}}},{\lambda} ^{{\mathcal t}},{\lambda ^{{\mathcal r}}}\hfill} \right\}} {{\rm{{\cal F}}}}\left( {{\bm z},{{\bm v}^{\mathcal t}},{{\bm v}^{\mathcal r}},{{\lambda}^{\mathcal t}},{{\lambda}^{\mathcal r}}} \right) + \frac{\gamma  }{2}\sum\limits_{{\mathcal p} = {\mathcal t},{\mathcal r}} {\left| {{{\bm v}^{\mathcal p}} - {{\bm \varphi} ^{\mathcal p}}} \right|_2^2} \label{eq:5a}\\
		&\;\;\;\;\;\;\;\;s.t.\;\;\;\left\{ \begin{array}{l}\vspace{1ex}
			\left| {\varphi_m^{\mathcal t}} \right| = \left| {\varphi_m^{\mathcal r}} \right| =1,\left\{ {{{\lambda} ^{{\mathcal t}}},{{\lambda} ^{{\mathcal r}}}} \right\}\ge 0,\quad\;\;\;\;\;{\rm if\;TS}\\\vspace{1ex}
			\left| {\varphi_m^{\mathcal t}} \right|,\left| {\varphi_m^{\mathcal r}} \right| \in \left\{ {0,1} \right\},{{\lambda} ^{{\mathcal t}}}={{\lambda} ^{{\mathcal r}}}=1,\;\;\;\;\;{\rm if\;MS}\\
			\begin{array}{l}
				\angle \varphi_m^{\mathcal t} - \angle \varphi_m^{\mathcal r} \in \left\{ {{\pi  \mathord{\left/
							{\vphantom {\pi  2}} \right.
							\kern-\nulldelimiterspace} 2}\left( {\bmod 2\pi } \right),} \right.\\
				\left. { - {\pi  \mathord{\left/
							{\vphantom {\pi  2}} \right.
							\kern-\nulldelimiterspace} 2}\left( {\bmod 2\pi } \right)} \right\},{\lambda ^{\mathcal t}} = {\lambda ^{\mathcal r}} = 1,
			\end{array}\;\;\;\;\;\;{\rm if\;ES}
		\end{array} \right.\label{eq:5c}\\
		&\;\;\;\;\;\;\;\;\;\;\;\;\;\;\;\;\;\angle \varphi_m^{\mathcal t},\angle \varphi_m^{\mathcal r} \in \left\{ {0,{{2\pi } \mathord{\left/
					{\vphantom {{2\pi } L}} \right.
					\kern-\nulldelimiterspace} L}, \cdots ,{{2\pi \left( {L - 1} \right)} \mathord{\left/
					{\vphantom {{2\pi \left( {L - 1} \right)} L}} \right.
					\kern-\nulldelimiterspace} L}} \right\},\label{eq:5d}\\
				&\;\;\;\;\;\;\;\;\;\;\;\;\;\;\;\;\;\eqref{eq:4b}, \eqref{eq:4e} \nonumber
	\end{align}
\end{subequations}
where $\gamma $ is the penalty coefficient. When the penalty coefficient increases, the RIS coefficient vectors $\bm v^{\mathcal t}$ and $\bm v^{\mathcal r}$ will be forced to take the same values as the auxiliary vectors $\bm \varphi^{\mathcal t}$ and $\bm \varphi^{\mathcal r}$, respectively, which makes $\bm v^{\mathcal t}$ and $\bm v^{\mathcal r}$ satisfy the constraints \eqref{eq:4c} and \eqref{eq:4d}. 

Recognizing that the constraints for $\left\{ {{{\bm \varphi} ^{\mathcal t}},{{\bm \varphi} ^{\mathcal r}}} \right\}$ and $\left\{ {{\bm z},{{\bm v}^{\mathcal t}},{{\bm v}^{\mathcal r}},{{\lambda}^{\mathcal t}},{{\lambda}^{\mathcal r}}} \right\}$ in \eqref{eq:5} are not coupled, BCD framework can be adopted to handle this problem, which involves the alternatively solving of the following two subproblems:
\begin{equation*}
	\begin{split}
		{\rm P1}: \mathop {\min }\limits_{\left\{ {{{\bm \varphi} ^{\mathcal t}},{{\bm \varphi} ^{\mathcal r}}} \right\}} \;\;& {\left| {{{\bm v}^{\mathcal t}} - {{\bm \varphi} ^{\mathcal t}}} \right|_2^2 + \left| {{{\bm v}^{\mathcal r}} - {{\bm \varphi} ^{\mathcal r}}} \right|_2^2} \\
		s.t. \;\;\;\;&\left\{ \begin{array}{l}\vspace{1ex}
			\left| {\varphi_m^{\mathcal t}} \right| = \left| {\varphi_m^{\mathcal r}} \right| =1,\qquad\qquad\qquad\;\;\;\;\;\;{\rm if\;TS}\\ \vspace{1ex}
			\left| {\varphi_m^{\mathcal t}} \right|,\left| {\varphi_m^{\mathcal r}} \right| \in \left\{ {0,1} \right\},\;\;\;\;\;\;\;\;\;\;\;\;\;\;\;\;\;\;\;\;\;\;\;{\rm if\;MS}\\
			\begin{array}{l}
				\angle \varphi_m^{\mathcal t} - \angle \varphi_m^{\mathcal r} \in \left\{ {{\pi  \mathord{\left/
							{\vphantom {\pi  2}} \right.
							\kern-\nulldelimiterspace} 2}\left( {\bmod 2\pi } \right),} \right.\\
				\left. { - {\pi  \mathord{\left/
							{\vphantom {\pi  2}} \right.
							\kern-\nulldelimiterspace} 2}\left( {\bmod 2\pi } \right)} \right\},
			\end{array}
			\;\;\;{\rm if\;ES}
		\end{array} \right.\\
		&\;\angle \varphi_m^{\mathcal t},\angle \varphi_m^{\mathcal r} \in \left\{ {0,{{2\pi } \mathord{\left/
					{\vphantom {{2\pi } L}} \right.
					\kern-\nulldelimiterspace} L}, \cdots ,{{2\pi \left( {L - 1} \right)} \mathord{\left/
					{\vphantom {{2\pi \left( {L - 1} \right)} L}} \right.
					\kern-\nulldelimiterspace} L}} \right\}.
			\end{split}
	\end{equation*}
\begin{equation*}
	\begin{split}
		{\rm P2}: \mathop {\min }\limits_{\left\{ {\scriptstyle {\bm z},{{\bm v}^{{\mathcal t}}},{{\bm v}^{{\mathcal r}}},\hfill\atop
				\scriptstyle{{\lambda} ^{{\mathcal t}}},{\lambda ^{{\mathcal r}}}\hfill} \right\}}& {\cal F}\left( {{\bm z},{{\bm v}^{\mathcal t}},{{\bm v}^{\mathcal r}},{\lambda ^{\mathcal t}},{\lambda ^{\mathcal r}}} \right) + \frac{\gamma }{2}\sum\limits_{{\mathcal p} = {\mathcal t},{\mathcal r}} {\left| {{{\bm v}^{\mathcal p}} - {{\bm \varphi} ^{\mathcal p}}} \right|_2^2}\\
		s.t.\;\;\;\;\;&\left\{ \begin{array}{l}\vspace{1ex}
			{\lambda ^{\mathcal t}} + {\lambda ^{\mathcal r}} = 1,\left\{ {{\lambda ^{\mathcal t}},{\lambda ^{\mathcal r}}} \right\} \ge 0,\;\;\;\;\;{\rm if\;TS}\\ 
			\begin{array}{l}
				{\left| {v_m^{\mathcal t}} \right|^2} + {\left| {v_m^{\mathcal r}} \right|^2} = 1,\\
				{\lambda ^{\mathcal t}} = {\lambda ^{\mathcal r}} = 1,
			\end{array}\quad\;\;\;\;\;\;\;\;\;\;\;{\rm if\;MS/ES}
		\end{array} \right.\\
		&\left( {{\bm z},{{\bm v}^{\mathcal t}},{{\bm v}^{\mathcal r}},{\lambda ^{\mathcal t}},{\lambda ^{\mathcal r}}} \right) \in {\bm \Omega }
	\end{split}
\end{equation*}
The advantage of solving subproblems P1 and P2 iteratively is that it separates the constraints with discrete selection (i.e., STAR-RIS coefficients of \eqref{eq:5c} and \eqref{eq:5d}) from the objective function ${\rm{{\cal F}}}$. This means that when we solve subproblem P2, we do not need to consider discrete selection constraints. Together with an increasing weight of the penalty term, the overall algorithm for solving \eqref{eq:4} is summarized in \textbf{Algorithm 1}. 

In a recent work dealing with ES STAR-RIS with coupled-phase~\cite{Wang:23}, the idea of penalty is also employed. The difference in the proposed framework is that we cover different STAR-RIS types (shown in Table I) while~\cite{Wang:23} is only tailored for ES STAR-RIS with coupled-phase. Furthermore, we include discrete phase constraint in the penalty while~\cite{Wang:23} did not, making the proposed framework more general. More importantly, we retain the constraint \eqref{eq:4b} in the subproblem P2 with respect to the original optimization variables, while~\cite{Wang:23} enforces \eqref{eq:4b} using auxiliary variables. Although the last point seems to be a unremarkable difference, this subtle change leads to significant consequences due to the following reasons:

\begin{enumerate}[leftmargin=0.5cm]
	
	\item For the subproblem P1, closed-form global optimal solutions for all STAR-RIS models in Table I can be obtained, even with the presence of discrete phase constraint \eqref{eq:4d}. Details will be presented in the next section. In contrast, in the corresponding subproblem of~\cite{Wang:23}, it requires alternatively solving for the amplitude and phase of ${{{\bm \varphi} ^{\mathcal t}}}$ and ${{{\bm \varphi} ^{\mathcal r}}}$, which slow down the convergence of the algorithm at the subproblem level.
	
	\item Under the special case of ES STAR-RIS with coupled-phase in \eqref{eq:5}, the split of the constraints \eqref{eq:4b} and \eqref{eq:5c}-\eqref{eq:5d} means that we are moving the solution of basic ES STAR-RIS (corresponding to subproblem P2) toward the solution of the coupled-phase ES STAR-RIS. However, the penalty method in~\cite{Wang:23} is to move the unconstrained STAR-RIS (with no operating mode information) solution to approach the coupled-phase ES STAR-RIS solution. In this case, the penalty weight required by the framework in~\cite{Wang:23} would be larger than that of the proposed framework to reach convergence. Since the penalty loop is the outer layer of the algorithm, a small penalty ratio for reaching convergence reduces the computational time of the proposed algorithm. This point will be further illustrated in the simulation results section.
	
\end{enumerate}

\renewcommand{\algorithmicrequire}{\textbf{Input:}} 
\renewcommand{\algorithmicensure}{\textbf{General step:}} 
\begin{algorithm}[t]
	\caption{Penalty-based BCD algorithm}
	\begin{algorithmic}[1]
		\REQUIRE ~~\\
		
		Initialize $\gamma$, increasing ratio of the penalty $c>1$, and the penalty fulfillment threshold $\delta$.
		\ENSURE ~~\\
		Initialize a feasible starting point for $\bm z$, ${{\bm v}^{\mathcal t}}$, ${{\bm v}^{\mathcal r}}$, ${{\bm \varphi} ^{\mathcal t}}$, ${{\bm \varphi} ^{\mathcal r}}$, ${{\lambda} ^{\mathcal t}}$, ${{\lambda} ^{\mathcal r}}$ .\\
		\textbf{Do}\\
		\quad \textbf{For} $n = 0,1,2,...$ execute the following steps:\\
		\quad \quad optimize $\left\{ {{\bm z},{{\bm v}^{\mathcal t}},{{\bm v}^{\mathcal r}},{{\lambda}^{\mathcal t}},{{\lambda}^{\mathcal r}}} \right\}$ by solving P2.\\
		\quad \quad optimize $\left\{ {{{\bm \varphi} ^{\mathcal t}},{{\bm \varphi} ^{\mathcal r}}} \right\}$ by solving P1.\\
		\quad \textbf{until} the objective function \eqref{eq:5a} converge.\\
		\quad $\gamma  \leftarrow c\gamma $.\\
		\textbf{until} $\mathop {\max }\limits_m \left| {v_m^{\mathcal t} - \varphi _m^{\mathcal t}} \right| \le \delta$ and $\mathop {\max }\limits_m \left| {v_m^{\mathcal r} - \varphi _m^{\mathcal r}} \right| \le \delta$.\\
	\end{algorithmic}
\end{algorithm}

Before we present the solution to the subproblem P1 in the next section, we reveal the following proposition about the solution quality of the \textbf{Algorithm 1}.

\begin{myPro}
	Suppose the solution from solving ${\rm P2}$ takes finite value for all $n$. Also assume that a stationary solution of ${\rm P2}$ and the global optimal solution of ${\rm P1}$ can be obtained. Then we have
	
\begin{enumerate}[leftmargin=0.5cm]
	
	\item The limit point generated by the BCD loop in \textbf{Algorithm 1} is a stationary solution of \eqref{eq:5} under any fixed penalty $\gamma$.
	
	\item As the penalty weight $\gamma$ increase to $\infty $, the limit point generated by \textbf{Algorithm 1} is a stationary point of the original problem \eqref{eq:4}.
	
\end{enumerate}
\end{myPro}
\noindent Proof. Please refer to Appendix A.

\begin{figure} 
	\centering
	\includegraphics[width=1\linewidth]{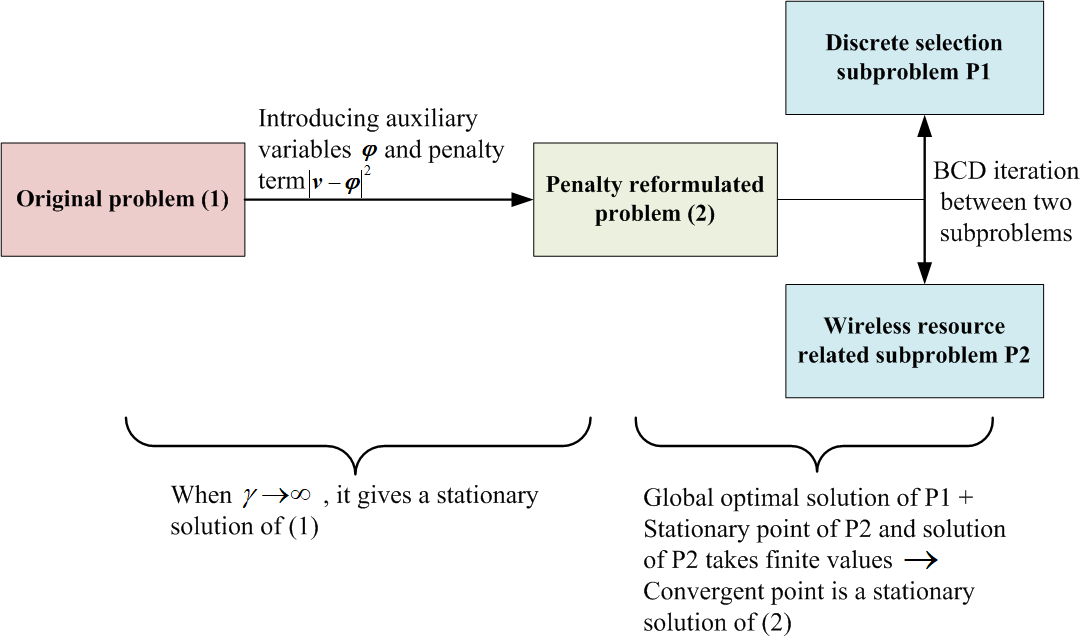}
	\caption{Summary of the proposed framework and its solution quality}
	\label{fig:graph}
\end{figure}
\textbf{Proposition 1} gives the legitimacy of using BCD under a fixed penalty parameter, and the penalty framework to enforce the STAR-RIS constraints in \eqref{eq:4}. Notice that convergence of solution also guarantees the convergence of the objective function value. Therefore, \textbf{Proposition 1} is stronger than just the objective function value convergence. The first assumption of \textbf{Proposition 1} is a mild assumption as solution with infinite value is useless in practical applications. For the requirement of the stationary solution of P2, since no discrete selection constraint is involved, it is easily satisfied from solutions obtained by successive convex approximation (SCA) technique~\cite{Bedi:22,Li:19} or first-order optimization methods~\cite{beck:17,Zong:21} in many communication problems. On the other hand, it seems quite daunting at the first sight as the global optimal solution of P1 is required. We will reveal in the next section that this is possible. To facilitate the understanding of the proposed penalty framework and the solution quality guarantee by \textbf{Proposition 1}, Fig. 3 summarizes various key problem formulations and the conditions of convergence.
\section{optimizing auxiliary variables $\bm \varphi $ in P1}
Notice that the elements of ${{\bm \varphi} ^{\mathcal t}}$ and ${{\bm \varphi} ^{\mathcal r}}$ are not coupled in P1. Therefore, P1 can be parallelized into $M$ subproblems with the $m^{th}$ subproblem given by

\begin{equation}\label{eq:6}
	\begin{split}
		\mathop {\min }\limits_{\varphi _m^{\mathcal t},\varphi _m^{\mathcal r}} \;\;&\left| {v_m^{\mathcal t} - \varphi _m^{\mathcal t}} \right|^2 + \left| {v_m^{\mathcal r} - \varphi _m^{\mathcal r}} \right|^2\\
		s.t. \;\;\;\;&\left\{ \begin{array}{l}\vspace{1ex}
			\left| {\varphi_m^{\mathcal t}} \right| = \left| {\varphi_m^{\mathcal r}} \right| =1,\;\;\;\;\;\;\;\;\;\;\;\;\;\;\;\;\;\;\;\;\;\;\;\;\;\;\;{\rm if\;TS}	\\\vspace{1ex}
			\left| {\varphi_m^{\mathcal t}} \right|,\left| {\varphi_m^{\mathcal r}} \right| \in \left\{ {0,1} \right\},\;\;\;\;\;\;\;\;\;\;\;\;\;\;\;\;\quad\;\;\; {\rm if\;MS}\\
			\begin{array}{l}
				\angle \varphi_m^{\mathcal t} - \angle \varphi_m^{\mathcal r} \in \left\{ {{\pi  \mathord{\left/
							{\vphantom {\pi  2}} \right.
							\kern-\nulldelimiterspace} 2}\left( {\bmod 2\pi } \right),} \right.\\
				\left. { - {\pi  \mathord{\left/
							{\vphantom {\pi  2}} \right.
							\kern-\nulldelimiterspace} 2}\left( {\bmod 2\pi } \right)} \right\},
			\end{array}\;\;\;{\rm if\;ES}
		\end{array} \right.\\
		&\;\angle \varphi_m^{\mathcal t},\angle \varphi_m^{\mathcal r} \in \left\{ {0,{{2\pi } \mathord{\left/
					{\vphantom {{2\pi } L}} \right.
					\kern-\nulldelimiterspace} L}, \cdots ,{{2\pi \left( {L - 1} \right)} \mathord{\left/
					{\vphantom {{2\pi \left( {L - 1} \right)} L}} \right.
					\kern-\nulldelimiterspace} L}} \right\}.\\
	\end{split}
\end{equation}
As shown in Table I, different combinations of the constraints in \eqref{eq:6} result in different types of STAR-RIS. Below, we divide the discussion in two cases. First, we consider the cases without coupled-phase constraint, which corresponds to cases 1-6 in Table I. Then we discuss coupled-phase cases 7 and 8 in Table I.

$\;$\underline{\emph{STAR-RISs without coupled-phase:}} The resulting subproblem is 
\begin{subequations}\label{eq:7}
	\begin{align}
		\mathop {\min }\limits_{\varphi _m^{\mathcal t},\varphi _m^{\mathcal r}}\;\;& \left| {v_m^{\mathcal t} - \varphi _m^{\mathcal t}} \right|^2 + \left| {v_m^{\mathcal r} - \varphi _m^{\mathcal r}} \right|^2\label{eq:7a}\\
		s.t. \;\;\;\;&\left\{ \begin{array}{l}\vspace{1ex}
			\left| {\varphi_m^{\mathcal t}} \right|=\left| {\varphi_m^{\mathcal r}} \right| =1,\;\;\;\;\;\;\;\;\;\;\;\;\;\;\;\;\;\;\;\;\;\;\;\;\;{\rm if\;TS}\\\vspace{1ex}
			\left| {\varphi_m^{\mathcal t}} \right|,\left| {\varphi_m^{\mathcal r}} \right|  = \left\{ {0,1} \right\}{\rm \;or\;}\left\{ {1,0} \right\},\;\;\;\;\; {\rm if\;MS}\\
			\varphi _m^{\mathcal t},\varphi _m^{\mathcal r} \in {\mathbb C},\;\;\;\;\;\;\;\;\;\;\;\;\;\;\;\;\;\;\;\;\;\;\;\;\;\;\;\;\;\;\;\;{\rm if\;ES}	
		\end{array} \right.\label{eq:7b}\\
		&\;\angle \varphi _m^{\mathcal t},\angle \varphi _m^{\mathcal r} \in \left\{ {0,{{2\pi } \mathord{\left/
					{\vphantom {{2\pi } L}} \right.
					\kern-\nulldelimiterspace} L}, \cdots ,{{2\pi \left( {L - 1} \right)} \mathord{\left/
					{\vphantom {{2\pi \left( {L - 1} \right)} L}} \right.
					\kern-\nulldelimiterspace} L}} \right\}.\label{eq:7c}
	\end{align}
\end{subequations}
Generally, this is a problem containing interger variables since the phase is discrete, which is hard to solve. However, since the amplitude and phase constraints are not coupled in \eqref{eq:7}, we can solve them separately and obtain closed-form solution of \eqref{eq:7} given by the following lemma.

\begin{myLem}
	Define $\alpha _m^{\mathcal t} = {{\rm Proj}_{{\bm \Theta}}}\left( {\angle v_m^{\mathcal t}} \right)$, $\alpha _m^{\mathcal r} = {{\rm Proj}_{{\bm \Theta }}}\left( {\angle v_m^{\mathcal r}} \right)$, $\beta _m^{\mathcal t} = \left| {v_m^{\mathcal t}} \right|\cos \left( {\alpha _m^{\mathcal t} - \angle v_m^{\mathcal t}} \right)$ and $\beta _m^{\mathcal r} = \left| {v_m^{\mathcal r}} \right|\cos \left( {\alpha _m^{\mathcal r} - \angle v_m^{\mathcal r}} \right)$. The optimal solution of \eqref{eq:7} is 
	\begin{equation}\label{eq:8}
		 \begin{array}{l}\vspace{1ex}
			\varphi _m^{\mathcal t} = {e^{j\alpha _m^{\mathcal t}}},\varphi _m^{\mathcal r} = {e^{j\alpha _m^{\mathcal r}}}, \;\;\;\;\;\;\;\;\;\;\;\;\;\;\;\;\;\;\;\;\;\;\;\;\;\;\;\;\;\;\;\;\;\;\;\;\;\;\;\;\;\;\;\;{\rm if\; TS}\\\vspace{1ex}
			\varphi _m^{\mathcal t} = \frac{{1 + {\mathop{\rm sgn}} \left( {\beta _m^{\mathcal t} - \beta _m^{\mathcal r}} \right)}}{2}{e^{j\alpha _m^{\mathcal t}}},\varphi _m^{\mathcal r} = \frac{{1 + {\mathop{\rm sgn}} \left( {\beta _m^{\mathcal r} - \beta _m^{\mathcal t}} \right)}}{2}{e^{j\alpha _m^{\mathcal r}}},{\rm if\; MS}\\
			\varphi _m^{\mathcal t} = \beta _m^{\mathcal t}{e^{j\alpha _m^{\mathcal t}}},\varphi _m^{\mathcal r} = \beta _m^{\mathcal r}{e^{j\alpha _m^{\mathcal r}}} \;\;\;\;\;\;\;\;\;\;\;\;\;\;\;\;\;\;\;\;\;\;\;\;\;\;\;\;\;\;,\;\;\;\;{\rm if\; ES}
		\end{array}
	\end{equation}
where ${\bm \Theta } = \left\{ {0,{{2\pi } \mathord{\left/
			{\vphantom {{2\pi } {L}}} \right.
			\kern-\nulldelimiterspace} L}, \cdots ,{{2\pi \left( {L - 1} \right)} \mathord{\left/
			{\vphantom {{2\pi \left( {L - 1} \right)} L}} \right.
			\kern-\nulldelimiterspace} L}} \right\}$, ${{\rm Proj}_{\rm{{\cal A}}}}\left( b \right)$ is to project $b$ to the set ${\rm{{\cal A}}}$, and ${\mathop{\rm sgn}} $ is the sign function. 
\end{myLem}
\noindent Proof. Please refer to Appendix B.

$\;$\underline{\emph{STAR-RISs with coupled-phase:}} Noting that coupled-phase only occurs in ES mode, leading to the subproblem P1 reduces to 
\begin{subequations}\label{eq:9}
	\begin{align}
		\mathop {\min }\limits_{\varphi _m^{\mathcal t},\varphi _m^{\mathcal r}} \;\;&\left| {v_m^{\mathcal t} - \varphi _m^{\mathcal t}} \right|^2 + \left| {v_m^{\mathcal r} - \varphi _m^{\mathcal r}} \right|^2\label{eq:9a}\\
		s.t.\;\;\;\;&\angle \varphi_m^{\mathcal t} - \angle \varphi_m^{\mathcal r} \in \left\{ {{\pi  \mathord{\left/
					{\vphantom {\pi  2}} \right.
					\kern-\nulldelimiterspace} 2}\left( {\bmod 2\pi } \right),{{ - \pi } \mathord{\left/
					{\vphantom {{ - \pi } 2}} \right.
					\kern-\nulldelimiterspace} 2}\left( {\bmod 2\pi } \right)} \right\}
		,\label{eq:9b}\\
		&\angle \varphi _m^{\mathcal t},\angle \varphi _m^{\mathcal r} \in \left\{ {0,{{2\pi } \mathord{\left/
					{\vphantom {{2\pi } L}} \right.
					\kern-\nulldelimiterspace} L}, \cdots ,{{2\pi \left( {L - 1} \right)} \mathord{\left/
					{\vphantom {{2\pi \left( {L - 1} \right)} L}} \right.
					\kern-\nulldelimiterspace} L}} \right\}.\label{eq:9c}
	\end{align}
\end{subequations}
As illustrated below \eqref{eq:4}, when $L>2$ and is an even number, constraint ~\eqref{eq:9b} and \eqref{eq:9c} are compatible. In this case, the global optimal solution of \eqref{eq:9} is given in \textbf{Lemma 2}.
\begin{myLem} If $L>2$ and is an even number, the optimal solution of \eqref{eq:9} is 
	\begin{equation}\label{eq:10}
		 \begin{array}{l}
			\varphi _m^{\mathcal t} = \left| {v_m^{\mathcal t}} \right|\cos \left( {\theta _m^{\mathcal t} - \angle v_m^{\mathcal t}} \right){e^{j\theta _m^{\mathcal t}}},\\
			\varphi _m^{\mathcal r} = \left| {v_m^{\mathcal r}} \right|\left| {\sin \left( {\theta _m^{\mathcal t} - \angle v_m^{\mathcal r}} \right)} \right|{e^{j\left( {\theta _m^{\mathcal t} - \frac{\pi }{2}{\mathop{\rm sgn}} \left( {\sin \left( {\theta _m^{\mathcal t} - \angle v_m^{\mathcal r}} \right)} \right)} \right)}},
		\end{array} 
	\end{equation}
where $\theta _m^{{\mathcal t}} = {{\rm Proj}_{{\bm \Theta }}}\left( {\angle v_m^{\mathcal t} {{ - {b_m}} \mathord{\left/
			{\vphantom {{ - {b_m}} 2}} \right.
			\kern-\nulldelimiterspace} 2} + {\pi  \mathord{\left/
			{\vphantom {\pi  2}} \right.
			\kern-\nulldelimiterspace} 2}} \right)$ and
${b_m} =  - j\ln $ $\left( {{\frac{{\left[ {{{\left| {v_m^{\mathcal r}} \right|}^2}\cos \left( {2\angle v_m^{\mathcal t} - 2\angle v_m^{\mathcal r}} \right) + {{\left| {v_m^{\mathcal t}} \right|}^2}} \right] + j{{\left| {v_m^{\mathcal r}} \right|}^2}\sin \left( {2\angle v_m^{\mathcal t} - 2\angle v_m^{\mathcal r}} \right)}}{{\sqrt {{{\left[ {{{\left| {v_m^{\mathcal r}} \right|}^2}\cos \left( {2\angle v_m^{\mathcal t} - 2\angle v_m^{\mathcal r}} \right) + {{\left| {v_m^{\mathcal t}} \right|}^2}} \right]}^2} + {{\left[ {{{\left| {v_m^{\mathcal r}} \right|}^2}\sin \left( {2\angle v_m^{\mathcal t} - 2\angle v_m^{\mathcal r}} \right)} \right]}^2}} }}}} \right)$.
\end{myLem}

\noindent Proof. Please refer to Appendix C.

\textbf{Lemmas 1} and \textbf{2} cover both discrete and continuous phase STAR-RIS. For the latter case, it is equivalent to taking $L \to \infty $, and the projection functions in \textbf{Lemmas 1} and \textbf{2} can be skipped. It is worth noting that for the simplest ES STAR-RIS (the fifth case in Table I), since there is no phase and amplitude constraint involved in \eqref{eq:6}, the optimal solution of the auxiliary variables ${\bm \varphi ^{\mathcal t}}$ and ${\bm \varphi ^{\mathcal r}}$ will always be equal to ${\bm v ^{\mathcal t}}$ and ${\bm v ^{\mathcal r}}$, respectively. Hence, the penalty loop would only be executed once, which reduces to the conventional non-penalty design in many existing works~\cite{Zhang:22b,Hou:22}.

\section{A case study on Downlink STAR-RIS assisted transmission system}
As an example for the proposed framework, we consider a downlink STAR-RIS assisted communication system. Assume that the BS has $N$ antennas and the STAR-RIS comprises $M$ elements. Furthermore, there are $K^{\mathcal r}$ users and $K^{\mathcal t}$ users with single antenna in the reflection region and the transmission region, respectively. In this paper, we assume that channel state information (CSI) is available at the BS. Let $s_i \in {{\rm{{\mathbb C}}}}$ with normalized power represents the information symbol targeted to the $i^{th}$ user ($i = 1,...,K^{\mathcal r} + K^{\mathcal t}$). With the individual beamforming vectors ${{\bm w}_i} \in {{\rm{{\mathbb C}}}^{N \times 1}}$ applied at the BS, the signal transmitted from the BS is $\Sigma_{i = 1}^{K^{\mathcal r} + K^{\mathcal t}} {{{\bm w}_i}{s_i}} $. Define the reflection user set and transmission user set as ${{\rm{{\cal K}}}^{\mathcal r}} = \left\{ {1, \cdots ,{K^{\mathcal r}}} \right\}$ and ${{\rm{{\cal K}}}^{\mathcal t}} = \left\{ {{K^{\mathcal r}}+1, \cdots ,{K^{\mathcal r}}+{K^{\mathcal t}}} \right\}$ respectively,
the received signal of the $l^{th}$ user is
\begin{equation}\label{eq:11}
{y_l} = \left\{ \begin{array}{l}
	\sum\limits_{i = 1}^{K^{\mathcal r} + K^{\mathcal t}} {\left( {{\bm h}_l^T{\rm diag}\left( {{{\bm v}^{\mathcal r}}} \right){\bm G} + {\bm d}_l^T} \right){{\bm w}_i}{s_i}}  + {n_l},\;\;l \in {{\rm{{\cal K}}}^{\mathcal r}}\\
	\sum\limits_{i = 1}^{K^{\mathcal r} + K^{\mathcal t}} {\left( {{\bm h}_l^T{\rm diag}\left( {{{\bm v}^{\mathcal t}}} \right){\bm G} + {\bm d}_l^T} \right){{\bm w}_i}{s_i}}  + {n_l},\;\;l \in {{\rm{{\cal K}}}^{\mathcal t}}
\end{array} \right.,
\end{equation}
where ${\bm G} \in {{\rm{{\mathbb C}}}^{M \times N}}$ and ${{\bm h}_{l}} \in {{\rm{{\mathbb C}}}^{M \times 1}}$ are the channels from the BS to the STAR-RIS and the STAR-RIS to the $l^{th}$ user, respectively. ${{\bm d}_{l}} \in {{\rm{{\mathbb C}}}^{N \times 1}}$ is the direct link channel from the BS to the $l^{th}$ user and ${n_{l}} \in {{\rm{{\mathbb C}}}}$ is Gaussian noise with zero mean and variance $\sigma _{l}^2$.

In this paper, we consider the downlink sum-rate maximization problem. Since the information symbols of different users and noise are uncorrelated, the sum-rate of the whole system is~\cite{Katwe:23}
\begin{equation}\label{eq:12}
	\begin{split}
		{\cal R}&\left( {{\bm w},{{\bm v}^{\mathcal t}},{{\bm v}^{\mathcal r}},{\lambda^{\mathcal t}},{\lambda^{\mathcal r}}} \right) =\\
		&{\lambda^{\mathcal r}}\sum\limits_{l = 1}^{{K^{\mathcal r}}} {\log \left( {1 + \frac{{{{\left| {{\bm a}_l^T{{\bm w}_l}} \right|}^2}}}{{\sum\nolimits_{i = 1, \ne l}^{{K^{\mathcal r}} + {K^{\mathcal t}}} {{{\left| {{\bm a}_l^T{{\bm w}_i}} \right|}^2}}  + {{\lambda}^{\mathcal r}}\sigma _l^2}}} \right)} \\
		&+ {\lambda^{\mathcal t}}\sum\limits_{l = {K^{\mathcal r}} + 1}^{{K^{\mathcal r}} + {K^{\mathcal t}}} {\log \left( {1 + \frac{{{{\left| {{\bm a}_l^T{{\bm w}_l}} \right|}^2}}}{{\sum\nolimits_{i = 1, \ne l}^{{K^{\mathcal r}} + {K^{\mathcal t}}} {{{\left| {{\bm a}_l^T{{\bm w}_i}} \right|}^2}}  + {{\lambda}^{\mathcal t}}\sigma _l^2}}} \right)},
	\end{split}
\end{equation}
where ${{\bm a}_l} = \left\{ \begin{array}{l}
	{{\bm G}^T}{\rm diag}\left( {{{\bm v}^{\mathcal r}}} \right){{\bm {h}}_l} + {{\bm {d}}_l},\;\;\;l \in {\cal K}^{\mathcal r}\\
	{{\bm G}^T}{\rm diag}\left( {{{\bm v}^{\mathcal t}}} \right){{\bm {h}}_l} + {{\bm {d}}_l},\;\;\;l \in {\cal K}^{\mathcal t}
\end{array} \right.$ and ${\bm w} = \left\{ {{{\bm w}_l}} \right\}_{l = 1}^{K^{\mathcal r} + K^{\mathcal t}}$. In \eqref{eq:12}, $\lambda^{\mathcal t}$ and $\lambda^{\mathcal r}$ are multiplied to the noise variance to ensure unit consistency with the interference term. This often occurs in system model involving time allocation~\cite{Lee:15,Ju:14}. Furthermore, for MS and ES modes, $\lambda^{\mathcal t}$ and $\lambda^{\mathcal r}$ are set to 1, and therefore \eqref{eq:12} reduces to traditional sum-rate expression.  

In \eqref{eq:12}, the set of beamforming vectors $\bm w$ and the RIS coefficients ${{\bm v}^{\mathcal t}}$, ${{\bm v}^{\mathcal r}}$ are nonlinearly coupled in both numerator and denominator. Besides, there is a summation of various data rates, which makes this objective function hard to tackle. If there is only one user, optimizing the data rate is equivalent to maximizing the signal-to-interference plus noise ratio (SINR) and hence quadratic transform~\cite{Shen:18} can be adopted to handle this single fraction. On the other hand, due to the presence of multiple users, we need the following equivalent sum-rate function for \eqref{eq:12}.
\begin{myLem}
	The sum-rate function ${\rm{{\cal R}}}\left( {{\bm w},{{\bm v} ^{{\mathcal t}}},{{\bm v} ^{{\mathcal r}}},{{\lambda} ^{{\mathcal t}}},{{\lambda} ^{{\mathcal r}}}} \right) = \mathop {\max }\limits_{{\bm \rho} ,{\bm x}} {{\rm{{\cal F}}}_1}\left( {{\bm w},{\bm \rho},{\bm x},{{\bm v} ^{\mathcal t}},{{\bm v} ^{\mathcal r}},{{\lambda} ^{\mathcal t}},{{\lambda} ^{\mathcal r}} } \right)$, where ${{\rm{{\cal F}}}_1}$ is shown in \eqref{eq:13} on the top of next page, ${\bm \rho}  = \left\{ {{\rho _1}, \cdots ,{\rho _{K^{\mathcal r} + K^{\mathcal t}}}} \right\}$ and ${\bm x} = \left\{ {{x_1}, \cdots ,{x_{K^{\mathcal r} + K^{\mathcal t}}}} \right\}$.
\end{myLem} 

\begin{figure*}[t]
	\begin{equation}\label{eq:13}
		\begin{split}
			{{\rm{{\cal F}}}_1}\left( {{\bm w},{\bm \rho} ,{\bm x},{{\bm v}^{\mathcal t}},{{\bm v}^{\mathcal r}},{{\lambda}^{\mathcal t}},{{\lambda}^{\mathcal r}}} \right) =& \lambda^{\mathcal r}\sum\limits_{l = 1}^{K^{\mathcal r}} {\left[ {\log \left( {1 + {\rho _l}} \right) - {\rho _l}} \right]} + \lambda^{\mathcal t}\sum\limits_{l = {K^{\mathcal r}}+1}^{K^{\mathcal r}+ K^{\mathcal t}} {\left[ {\log \left( {1 + {\rho _l}} \right) - {\rho _l}} \right]}\\
			&+ \lambda^{\mathcal r}\sum\limits_{l = 1}^{K^{\mathcal r}} {\left\{ {2\left( {1 + {\rho _l}} \right){\mathop{\rm Re}\nolimits} \left[ {\overline {{x_l}}  {{ {{\bm a}_l^T}}}{{\bm w}_l}} \right] - \left( {1 + {\rho _l}} \right){{\left| {{x_l}} \right|}^2}\left( {\sum\limits_{i = 1}^{K^{\mathcal r} + K^{\mathcal t}} {{{\left| {{{ {{\bm a}_l^T} }}{{\bm w}_i}} \right|}^2} + {{\lambda}^{\mathcal r}}\sigma _l^2} } \right)} \right\}}\\
			&+ \lambda^{\mathcal t}\sum\limits_{l = K^{\mathcal r}+1}^{K^{\mathcal r}+K^{\mathcal t}} {\left\{ {2\left( {1 + {\rho _l}} \right){\mathop{\rm Re}\nolimits} \left[ {\overline {{x_l}}  {{ {{\bm a}_l^T}}}{{\bm w}_l}} \right] - \left( {1 + {\rho _l}} \right){{\left| {{x_l}} \right|}^2}\left( {\sum\limits_{i = 1}^{K^{\mathcal r} + K^{\mathcal t}} {{{\left| {{{ {{\bm a}_l^T} }}{{\bm w}_i}} \right|}^2} + {{\lambda}^{\mathcal t}}\sigma _l^2} } \right)} \right\}},
		\end{split}
	\end{equation}
\rule[-10pt]{18.07cm}{0.1em}
\end{figure*} 

\noindent Proof. Please refer to Appendix D. 

Since the objective function is only involved in P2, we set the sum-rate function in \textbf{Lemma 3} as the objective of P2. Further recognizing that $\bm z$ in P2 corresponds to
$\left\{ {{\bm w},{\bm \rho}, {\bm x}} \right\}$ in this particular application, P2 can be written as
\setcounter{equation}{10}
\begin{equation}\label{eq:14}
	\begin{split}
	\mathop {\min }\limits_{\left\{ {\scriptstyle {\bm w},{\bm \rho} ,{\bm x},\hfill\atop
			{\scriptstyle{{\bm v}^t},{{\bm v}^r},\hfill\atop
				\scriptstyle{\lambda ^t},{\lambda ^r}\hfill}} \right\}} &-{\rm{{\cal F}}}_1\left( {{\bm w},{\bm \rho} ,{\bm x},{{\bm v} ^{{\mathcal t}}},{{\bm v} ^{{\mathcal r}}},{{\lambda} ^{{\mathcal t}}},{{\lambda} ^{{\mathcal r}}}} \right) + \frac{\gamma }{2}\sum\limits_{{\mathcal p} = {\mathcal t},{\mathcal r}}{{\left| {{{\bm v}^{\mathcal p}} - {{\bm \varphi} ^{\mathcal p}}} \right|_2^2}}\\
			s.t.\;\;\;\;&\left\{ \begin{array}{l}
			{\lambda^{\mathcal t}} + {\lambda^{\mathcal r}} = 1,\left\{ {{\lambda^{\mathcal t}},{\lambda^{\mathcal r}}} \right\} \ge 0,\;\;\;\;\;\; {\rm if\;TS}\\
			{\left| {v_m^{\mathcal t}} \right|^2} + {\left| {v_m^{\mathcal r}} \right|^2} = 1,\;\;\;\;\;\;\;\;\;\;\;\;\;\;\;\;\;\;\; {\rm if\;MS/ES}
		\end{array} \right.\\
		&\sum\limits_{l = 1}^{K^{\mathcal r} + K^{\mathcal t}} {\left| {{{\bm w}_l}} \right|_2^2 \le {P_{BS}}}.
	\end{split}
\end{equation}

Noticing that the constraints for ${\bm w}$, ${\bm \rho}$, ${\bm x}$, $\left\{ {{{\bm v}^{\mathcal t}},{{\bm v}^{\mathcal r}}} \right\}$ and $\left\{ {{{\lambda}^{\mathcal t}},{{\lambda}^{\mathcal r}}} \right\}$ are separable, BCD algorithm can be adopted to solve \eqref{eq:14}, and the details of solving different subproblems are given below. 
\subsection{Optimizing ${\bm x}$, ${\bm \rho}$, ${\bm w}$, ${{\lambda}^{\mathcal t}}$ and ${{\lambda}^{\mathcal r}}$}
Under the BCD formulation, the subproblems with respect to the auxiliary variables $\bm x$, $\bm \rho$, the beamforming vector $\bm w$ and time allocation variables $\left\{ {{\lambda^{\mathcal t}},{\lambda^{\mathcal r}}} \right\}$ are convex problems, and they are relatively easy to handle.

$\;$\underline{\emph{Optimizing auxiliary variable $\bm x$:}} Noting that ${{\rm{{\cal F}}}_1}$ is a convex function of $\bm x$ with each element $\left\{ {{x_l}} \right\}_{l = 1}^{K^{\mathcal r} + K^{\mathcal t}}$ being separable, we can take derivative of ${{\rm{{\cal F}}}_1}$ with respect to each $x_l$ and set them to zero, yielding the following the optimal solution
\begin{equation}\label{eq:15}
	{x_l} =	\left\{ \begin{array}{l}
		\frac{{{ {{\bm a}_l^T} }{{\bm w}_l}}}{{\Sigma_{i = 1}^{K^{\mathcal r} + K^{\mathcal t}} {{{\left| {{ {{\bm a}_l^T} }{{\bm w}_i}} \right|}^2} + \lambda^{\mathcal r}\sigma _l^2} }}, \;\;\;\;   l \in {\cal K}^{\mathcal r},
		\\
		\frac{{{ {{\bm a}_l^T} }{{\bm w}_l}}}{{\Sigma_{i = 1}^{K^{\mathcal r} + K^{\mathcal t}} {{{\left| {{ {{\bm a}_l^T} }{{\bm w}_i}} \right|}^2} + \lambda^{\mathcal t}\sigma _l^2} }}, \;\;\;\;   l \in {\cal K}^{\mathcal t}.
	\end{array} \right.
\end{equation}

$\;$\underline{\emph{Optimizing auxiliary variable $\bm \rho$:}} Since each element in $\bm \rho$ is also separable, by taking the derivative of ${{\rm{{\cal F}}}_1}$ with respect to each ${\rho _l}$ ($l = 1,2,...,K^{\mathcal r} + K^{\mathcal t}$) and set them to zero, we obtain
\begin{equation}\label{eq:16}
	{\rho _l} =\left\{ \begin{array}{l}
		\frac{{{{\left| {{ {{\bm a}_l^T} }{{\bm w}_l}} \right|}^2}}}{{\Sigma_{i = 1, \ne l}^{K^{\mathcal r} + K^{\mathcal t}} {{{\left| {{ {{\bm a}_l^T} }{{\bm w}_i}} \right|}^2} + \lambda^{\mathcal r}\sigma _l^2} }},  \;\;\;\;  l \in {\cal K}^{\mathcal r},
		\\
		\frac{{{{\left| {{ {{\bm a}_l^T} }{{\bm w}_l}} \right|}^2}}}{{\Sigma_{i = 1, \ne l}^{K^{\mathcal r} + K^{\mathcal t}} {{{\left| {{ {{\bm a}_l^T} }{{\bm w}_i}} \right|}^2} + \lambda^{\mathcal t}\sigma _l^2} }},  \;\;\;\;  l \in {\cal K}^{\mathcal t}.
	\end{array} \right.
\end{equation}

$\;$\underline{\emph{Optimizing beamforming vector $\bm w$:}} Focusing on the components related to $\bm w$ in \eqref{eq:14}, the following subproblem of $\bm w$ is obtained and it is a convex quadratic problem:
\begin{subequations}\label{eq:17}
	\begin{align}
		\mathop {\min }\limits_{\bm w} \;\;&\sum\limits_{l = 1}^{K^{\mathcal r} + K^{\mathcal t}} {\left\{ {{\bm w}_l^H{\bm \Xi}{{\bm w}_l} - 2{\mathop{\rm Re}\nolimits} \left[ {{\bm q}_l^H {{\bm w}_l}} \right]} \right\}} \label{eq:17a}\\
		s.t.\;\;&\sum\limits_{l = 1}^{K^{\mathcal r} + K^{\mathcal t}} {{\bm w}_l^H{{\bm w}_l}}  \le {P_{BS}},\label{eq:17b}
	\end{align}
\end{subequations}
where ${q_l} = \left\{ \begin{array}{l}
	{\lambda^{\mathcal r}}\left( {1 + {\rho _l}} \right){x_l}\overline {{{\bm a}_l}} ,\; l \in K^{\mathcal r} \\
	{\lambda^{\mathcal t}}\left( {1 + {\rho _l}} \right){x_l}\overline {{{\bm a}_l}} ,\; l \in K^{\mathcal t}
\end{array} \right.$ and ${\bm \Xi} = {\lambda^{\mathcal r}}\Sigma_{i = 1}^{K^{\mathcal r}} {\left( {1 + {\rho _i}} \right){{\left| {{x_i}} \right|}^2} \overline {{{\bm a}_i}}   { {\bm a}_i^T }} + {\lambda^{\mathcal t}}\Sigma_{i = K^{\mathcal r}+1}^{K^{\mathcal r}+K^{\mathcal t}} {\left( {1 + {\rho _i}} \right){{\left| {{x_i}} \right|}^2} \overline {{{\bm a}_i}}   { {\bm a}_i^T }}$. Notice that problem \eqref{eq:17} is a convex quadratically constrained quadratic program (QCQP), one popular option is to employ CVX to tackle this problem.
 
However, CVX tool does not take the advantage of the structure of \eqref{eq:17}. Since this problem has only one constraint, the following lemma with bisection method can be adopted to solve this problem optimally.
\begin{myLem}
	Denote the eigenvalue decomposition of $\bm \Xi$ as ${\bm U}{\bm  \Lambda }{{\bm U}^H}$, and ${\bm B} = \Sigma_{l = 1}^{K^{\mathcal r} + K^{\mathcal t}} {{{\bm U}^H}{{\bm q}_l}{\bm q}_l^H{\bm U}}$, the optimal beamforming vector of \eqref{eq:17} is ${{\bm w}_l} = {\bm U}{\left( {{\bm \Lambda}  + \mu {\bm I}} \right)^{ - 1}}{{\bm U}^H}{{\bm q}_l}$, where
	\begin{equation}\label{eq:18}
		\mu  = \left\{ \begin{array}{l}
			0, \qquad\qquad\qquad\qquad\qquad\;\;\;\;\;\;\;{\rm if}\;{\rm Tr}\left( {{{\bm \Lambda} ^{ - 2}}{\bm B}} \right) \le {P_{BS}}\\
			{\rm solution}\;{\rm of}\;{\rm Tr}\left( {{{\left( {{\bm \Lambda}  + \mu {\bm I}} \right)}^{ - 2}}{\bm B}} \right) = {P_{BS}},\;\;\;{\rm otherwise}
		\end{array} \right..
	\end{equation}
Since ${\rm Tr}\left( {{{\left( {{\bm \Lambda}  + \mu {\bm I}} \right)}^{ - 2}}{\bm B}} \right)$ is a strictly decreasing function of $\mu$, the solution of the second case in \eqref{eq:18} can be found by bisection method from interval $\left[ {0,\sqrt {{{{\rm Tr}\left( {\bm B} \right)} \mathord{\left/
				{\vphantom {{Tr\left( B \right)} {{P_{BS}}}}} \right.
				\kern-\nulldelimiterspace} {{P_{BS}}}}}  } \right]$.
\end{myLem}
\noindent Proof. Please refer to Appendix E.

$\;$\underline{\emph{Optimizing time allocation variables $\left\{ {{\lambda^{\mathcal t}},{\lambda^{\mathcal r}}} \right\}$:}} For ES and MS mode, we have ${\lambda^{\mathcal t}} = {\lambda^{\mathcal r}} = 1$. Hence, only ${{\lambda}^{\mathcal t}}$ and ${{\lambda}^{\mathcal r}}$ in the TS mode need to be optimized. Focusing on the components related to ${{\lambda}^{\mathcal t}}$ and ${{\lambda}^{\mathcal r}}$ in \eqref{eq:14}, the following subproblem is obtained,
\begin{subequations}\label{eq:108}
	\begin{align}
		\mathop {\min }\limits_{{\lambda ^{\mathcal t}},{\lambda ^{\mathcal r}}}\;\;& {\left( {{\lambda ^{\mathcal t}}} \right)^2}{\psi  ^{\mathcal t}} + {\left( {{\lambda ^{\mathcal r}}} \right)^2}{\psi ^{\mathcal r}} + {\lambda ^{\mathcal t}}\sum\limits_{l = {K^{\mathcal r}}+1}^{{K^{\mathcal r}}+K^{\mathcal t}} {{\eta _l^{\mathcal t}}}  + {\lambda ^{\mathcal r}}\sum\limits_{l = 1}^{{K^{\mathcal r}}} {{\eta _l^{\mathcal r}}}\label{eq:108a}\\
		s.t.\;\;&{\lambda^{\mathcal t}} + {\lambda^{\mathcal r}} = 1,\left\{ {{\lambda^{\mathcal t}},{\lambda^{\mathcal r}}} \right\} \ge 0,\label{eq:108b}
	\end{align}
\end{subequations}
where $\eta _l^{\mathcal r} = \left( {1 + {\rho _l}} \right)\left[ {{{\left| {{x_l}} \right|}^2}\sum\nolimits_{i = 1}^{{K^{\mathcal r}} + {K^{\mathcal t}}} {{{\left| {{\bm a}_l^T{{\bm w}_i}} \right|}^2} - 2{\mathop{\rm Re}\nolimits} \left( {\overline {{x_l}} {\bm a}_l^T{{\bm w}_l}} \right)} } \right]$ $- \log \left( {1 + {\rho _l}} \right) + {\rho _l}$, $\eta _l^{\mathcal t} = \left( {1 + {\rho _l}} \right)\left[ {{{\left| {{x_l}} \right|}^2}\sum\nolimits_{i = 1}^{{K^{\mathcal r}} + {K^{\mathcal t}}} {{{\left| {{\bm a}_l^T{{\bm w}_i}} \right|}^2}} } \right.-$ $\left. { 2{\mathop{\rm Re}\nolimits} \left( {\overline {{x_l}} {\bm a}_l^T{{\bm w}_l}} \right)} \right]- \log \left( {1 + {\rho _l}} \right) + {\rho _l}$, ${\psi ^{\mathcal r}} = \Sigma_{l = 1}^{{K^{\mathcal r}}} {\sigma _l^2}$ and ${\psi ^{\mathcal t}} = \Sigma_{l = {K^{\mathcal r}} + 1}^{{K^{\mathcal r}}+{K^{\mathcal t}}} {\sigma _l^2}$. Substituting ${{\lambda ^{\mathcal r}}}=1-{{\lambda ^{\mathcal t}}}$ in \eqref{eq:108a}, \eqref{eq:108} is equivalent to
\begin{equation}
	\begin{split}
		\mathop {\min }\limits_{{\lambda ^{\mathcal t}}}\;\;& \left( {{\psi ^{\mathcal t}} + {\psi ^{\mathcal r}}} \right){\left( {{\lambda ^{\mathcal t}} - \frac{{2{\psi ^{\mathcal r}} - {\eta ^{\mathcal t}} + {\eta ^{\mathcal r}}}}{{2{\psi ^{\mathcal t}} + 2{\psi ^{\mathcal r}}}}} \right)^2}\\
		s.t.\;\;&{\lambda ^{\mathcal t}} \in \left[ {0,1} \right],
	\end{split}
\end{equation}
where ${\eta ^{\mathcal t}} = \sum\nolimits_{l = {K^{\mathcal r}} + 1}^{{K^{\mathcal r}} + {K^{\mathcal t}}} {\eta _l^{\mathcal t}} $ and ${\eta ^{\mathcal r}} = \sum\nolimits_{l =  1}^{{K^{\mathcal r}}} {\eta _l^{\mathcal r}} $. Therefore, the optimal solution is ${\lambda ^{\mathcal t}} = {{\rm Proj}_{\left[ {0,1} \right]}}\left( {\frac{{2{\psi ^{\mathcal r}} - {\eta ^{\mathcal t}} + {\eta ^r}}}{{2{\psi ^{\mathcal t}} + 2{\psi ^{\mathcal r}}}}} \right)$ and ${\lambda ^{\mathcal r}} = 1-{{\rm Proj}_{\left[ {0,1} \right]}}\left( {\frac{{2{\psi ^{\mathcal r}} - {\eta ^{\mathcal t}} + {\eta ^{\mathcal r}}}}{{2{\psi ^{\mathcal t}} + 2{\psi ^{\mathcal r}}}}} \right)$.

\subsection{Optimizing ${{\bm v}^{\mathcal t}}$ and ${{\bm v}^{\mathcal r}}$}
To better illustrate the optimization subproblem with respect to ${{\bm v}^{\mathcal t}}$ and ${{\bm v}^{\mathcal r}}$, we define two shorthand notations:
\begin{equation}\label{eq:19}
 \begin{array}{l}
		{{\bm A}_l}=
		\\
		\left( {1 + {\rho _l}} \right){{\left| {{x_l}} \right|}^2}{\rm diag}\left( {{\bm h}_l^H} \right){\overline {\bm G} }\left( {\sum\limits_{i = 1}^{K^{\mathcal r} + K^{\mathcal t}} {{\overline {{{\bm w}_i}} }{\bm w}_i^T} } \right){{\bm G}^T}{\rm diag}\left( {{{\bm h}_l}} \right),
	\end{array}
\end{equation}

\begin{equation}\label{eq:20}
	\begin{array}{l}
		{{\bm b}_l} =
		\\
		2\left( {1 + {\rho _l}} \right){\rm diag}\left( {{\bm h}_l^H} \right){\overline {\bm G} }\left[ {{{\left| {{x_l}} \right|}^2}\left( {\sum\limits_{i = 1}^{K^{\mathcal r} + K^{\mathcal t}} {{\overline {{{\bm w}_i}} }{\bm w}_i^T} } \right){{\bm d}_l} -  x_l{\overline {{{\bm w}_l}} }} \right].\vspace{1ex}
	\end{array}
\end{equation}

Focusing on the terms related to ${\bm v}^{\mathcal t}$ and ${\bm v}^{\mathcal r}$ in \eqref{eq:14}, the subproblem becomes
\begin{subequations}\label{eq:21}
	\begin{align}
		\mathop {\min }\limits_{{{\bm v}^{\mathcal t}},{{\bm v}^{\mathcal r}}}&\sum\limits_{{\mathcal p} = {\mathcal t},{\mathcal r}} {{\left( {{{\bm v}^{\mathcal p}}} \right)^H}{{\bm A}^{\mathcal p}}{{\bm v}^{\mathcal p}} + {\mathop{\rm Re}\nolimits} \left[ {{{\left( {{{\bm b}^{\mathcal p}}} \right)}^H}{{\bm v}^{\mathcal p}}} \right]} \label{eq:21a}\\
		s.t.\;&{\left| {v_m^{\mathcal t}} \right|^2} + {\left| {v_m^{\mathcal r}} \right|^2} = 1,\;\; m \in \left\{ {1,2,...,M} \right\},\;\;\;\;\;{\rm if\;ES/MS}\label{eq:21b}
	\end{align}
\end{subequations}
where ${{\bm A}^{\mathcal r}} = {{\gamma {{\bm I}_M}} \mathord{\left/
		{\vphantom {{\lambda {{\bm I}_M}} 2}} \right.
		\kern-\nulldelimiterspace} 2}+{\lambda^{\mathcal r}}\Sigma_{l = 1}^{K^{\mathcal r}} {\bm A}_l$, ${{\bm A}^{\mathcal t}} = {{\gamma {{\bm I}_M}} \mathord{\left/
		{\vphantom {{\gamma {{\bm I}_M}} 2}} \right.
		\kern-\nulldelimiterspace} 2}+{\lambda^{\mathcal t}}\Sigma_{l = K^{\mathcal r} + 1}^{K^{\mathcal r} + K^{\mathcal t}} {\bm A}_l$,
${{\bm b}^{\mathcal r}} = - \gamma {{\bm \varphi} ^{\mathcal r}}+{\lambda^{\mathcal r}}\Sigma_{l = 1}^{K^{\mathcal r}} {{{\bm b}_l}}$ and ${{\bm b}^{\mathcal t}} = - \gamma {{\bm \varphi} ^{\mathcal t}} + {\lambda^{\mathcal t}}\Sigma_{l = K^{\mathcal r} + 1}^{K^{\mathcal r} + K^{\mathcal t}} {{{\bm b}_l}}  $.

For the TS mode, we do not have the constraint \eqref{eq:21b}. This is an unconstrained quadratic optimization and the closed-form solution is obtained by
\begin{equation}\label{eq:222}
	{{\bm v}^{\mathcal t}} =  - \frac{1}{2}{\left( {{{\bm A}^{\mathcal t}}} \right)^{ - 1}}{{\bm b}^{\mathcal t}}\;;\;{{\bm v}^{\mathcal r}} =  - \frac{1}{2}{\left( {{{\bm A}^{\mathcal r}}} \right)^{ - 1}}{{\bm b}^{\mathcal r}}.
\end{equation}
For the ES and MS modes, recognizing that \eqref{eq:21} has $M$ nonconvex and quadratic equality constraints, SCA is not suitable since the equality constraints cannot be approximated. On the other hand, noticing that the constraints \eqref{eq:21b} is separable with different index $m$, the coefficient pair $\left\{ {v_m^{\mathcal t},v_m^{\mathcal r}} \right\}$ can be sequentially updated under the BCD framework from $m = 1$ to $m = M$ with the rest coefficient pairs fixed. Hence, the subproblem with respect to $\left\{ {v_m^{\mathcal t},v_m^{\mathcal r}} \right\}$ is
\begin{equation}\label{eq:22}
	\begin{split}
		\mathop {\min }\limits_{v_m^{\mathcal t},v_m^{\mathcal r}} \;\;&\sum\limits_{{\mathcal p} = {\mathcal t},{\mathcal r}} {{\bm A}_{m,m}^{\mathcal p}{\left| {v_m^{\mathcal p}} \right|^2} + {\mathop{\rm Re}\nolimits} \left\{ { {{\left(\overline {c_m^{\mathcal p}} \right)}}v_m^{\mathcal p}} \right\}}\\
		s.t.\;\;\;&{\left| {v_m^{\mathcal t}} \right|^2} + {\left| {v_m^{\mathcal r}} \right|^2} = 1,
	\end{split}
\end{equation}
where $c_m^{{\mathcal t}} = b_m^{\mathcal t} + 2\Sigma_{j \ne m} {{\bm A}_{m,j}^{\mathcal t}v_j^{{\mathcal t}}}$ and $c_m^{{\mathcal r}} = b_m^{\mathcal r} + 2\Sigma_{j \ne m} {{\bm A}_{m,j}^{\mathcal r}v_j^{{\mathcal r}}}$, with ${\bm A}_{i,j}^{\mathcal t}$ and ${\bm A}_{i,j}^{\mathcal r}$ denote the ${\left( {i,j} \right)^{th}}$ element of ${{\bm A}^{\mathcal t}}$ and ${{\bm A}^{\mathcal r}}$, respectively. 

Expressing $v_m^{\mathcal t} = \left| {v_m^{\mathcal t}} \right|{e^{j\angle v _m^{\mathcal t}}}$ and $v_m^{\mathcal r} = \left| {v_m^{\mathcal r}} \right|{e^{j\angle v_m^{\mathcal r}}}$, and noticing that the phases ${\angle v _m^{\mathcal t}}$ and ${\angle v_m^{\mathcal r}}$ only affect the value of ${\mathop{\rm Re}\nolimits} \left\{ {{{\left( \overline {c_m^{\mathcal t}}  \right)}}v_m^{\mathcal t} + {{\left( \overline {c_m^{\mathcal r}}  \right)}}v_m^{\mathcal r}} \right\}$, the phases ${\angle v_m^{\mathcal t}}$ and ${\angle v_m^{\mathcal r}}$ that minimize \eqref{eq:22} should be chosen as
\begin{equation}\label{eq:23}
		\angle v_m^{\mathcal t} = \pi  + \angle c_m^{{\mathcal t}}\;\;\; ;
		\;\;\; \angle v_m^{\mathcal r} = \pi  + \angle c_m^{{\mathcal r}}.
\end{equation} 
Putting this result to \eqref{eq:22} and it reduces to
\begin{equation}\label{eq:24}
	\begin{split}
		\mathop {\min }\limits_{\left| {v_m^{\mathcal t}} \right|,\left| {v_m^{\mathcal r}} \right|}\;\;& \sum\limits_{{\mathcal p} = {\mathcal t},{\mathcal r}} {{\bm A}_{m,m}^{\mathcal p}{\left| {v_m^{\mathcal p}} \right|^2} -{\left| {c_m^{{\mathcal p}}} \right| \left| {v_m^{\mathcal p}} \right|} }\\
		s.t.\;\;\;\;&{\left| {v_m^{\mathcal t}} \right|^2} + {\left| {v_m^{\mathcal r}} \right|^2} = 1.
	\end{split}.
\end{equation}

Problem \eqref{eq:24} is a real-valued optimization problem with a circular constraint. This problem would be easier to solve if we define $\left| {v_m^{\mathcal r}} \right| = \cos \left( {{\phi _m}} \right)$ and $\left| {v_m^{\mathcal t}} \right| = \sin \left( {{\phi _m}} \right)$ and transfer the unknown to ${\phi _m} \in \left[ {0,{\pi  \mathord{\left/
			{\vphantom {\pi  2}} \right.
			\kern-\nulldelimiterspace} 2}} \right]$ with the constraint of \eqref{eq:24} guaranteed to satisfy. Then, the resulting one dimension unconstrained problem becomes
\begin{equation}\label{eq:25}
	\begin{split}
		\mathop {\min }\limits_{{\phi _m} \in \left[ {0,\frac{\pi }{2}} \right]} f\left( {{\phi _m}} \right) = &\left( {{\bm A}_{m,m}^{\mathcal r} - {\bm A}_{m,m}^{\mathcal t}} \right){\cos ^2}\left( {{\phi _m}} \right)\\
		&\;\;{- \left| c_m^{{\mathcal r}} \right|\cos \left( {{\phi _m}} \right) - \left| c_m^{{\mathcal t}} \right|\sin \left( {{\phi _m}} \right)}.
	\end{split}
\end{equation}
If we compute the gradient $\nabla f\left( {{\phi _m}} \right)$, it is found that 
\begin{equation}\label{eq:26}
	\begin{split}
	\nabla f\left( {{\phi _m}} \right) = &\sin \left( {{\phi _m}} \right)\cos \left( {{\phi _m}} \right)\\
	&\times\left[ {\frac{{\left| {c_m^{\mathcal r}} \right|}}{{\cos \left( {{\phi _m}} \right)}} - \frac{{\left| {c_m^{\mathcal t}} \right|}}{{\sin \left( {{\phi _m}} \right)}} - 2\left( {{\bm A}_{m,m}^{\mathcal r} - {\bm A}_{m,m}^{\mathcal t}} \right)} \right].
	\end{split}
\end{equation}
Since the function in the square bracket is a monotonic increasing function taking values from $\left( { - \infty ,\infty } \right)$, the zero point of the gradient can be obtained by the bisection method from the interval $\left[ {{{0,\pi } \mathord{\left/
			{\vphantom {{0,\pi } 2}} \right.
			\kern-\nulldelimiterspace} 2}} \right]$. Denoting the estimated $\phi _m$ of \eqref{eq:25} by bisection method as ${{\hat \phi }_m} $, the solution of \eqref{eq:22} is then given by
\begin{equation}\label{eq:27}
	v_m^{\mathcal t} = \sin \left( {{\hat \phi }_m} \right){e^{j\left( {\pi  + \angle c_m^{{\mathcal t}}} \right)}}\;\;\; ;\;\;\; v_m^{\mathcal r} = \cos \left( {{\hat \phi }_m} \right){e^{j\left( {\pi  + \angle c_m^{{\mathcal r}}} \right)}}.
\end{equation}
\subsection{Summary and Time Complexity of the Proposed Algorithm}
The proposed algorithms for sum-rate maximization under different STAR-RISs are summarized in Table II. From Table II, we can see the operating mode constraint and discrete phase constraint are handled using equations \eqref{eq:8} and \eqref{eq:10}. Hence, algorithm designers only need to focus on the subproblem P2. Since P2 under ES mode and MS mode are the same, there are only two variations (ES/MS and TS) for three kinds of STAR-RIS modes. Furthermore, from Table II, we can see that there are many common equations in algorithms for different STAR-RISs. This is how the proposed framework facilitates the design of algorithms under different types and operating modes of STAR-RIS.

\begin{table*}
	\centering
	\begin{threeparttable}[b]
		\caption{Equations related to different types of STAR-RIS}
		\label{tab:test2}
		\begin{tabular}{|c|c|c|c|c|c|}
			\hline
			& & & & &\\[-7pt]
			\tabincell{c}{Operating\\ mode}&\tabincell{c}{Coupled-phase\\constraint}&\tabincell{c}{Discrete phase\\constraint}&\tabincell{c}{Equations related\\to subproblem P1}&\tabincell{c}{Time allocation\\ constraint}&\tabincell{c}{Equations related\\to subproblem P2}\\
			\hline
			& & & & &\\[-7.5pt]
			\multirow{2}{*}{TS}&\usym{2718}&\usym{2718}&\multirow{2}{*}{First line of \eqref{eq:8}}&\CheckmarkBold&\multirow{2}{*}{\eqref{eq:15},\eqref{eq:16},\eqref{eq:18},\eqref{eq:222}}\\ \cline{2-3} \cline{5-5}
			& & & & &\\[-6pt]
			&\usym{2718}&\CheckmarkBold& &\CheckmarkBold&\\
			\hline
			& & & & &\\[-6pt]
			\multirow{2}{*}{MS}&\usym{2718}&\usym{2718}&\multirow{2}{*}{Second line of \eqref{eq:8}}&\usym{2718}&\multirow{6}{*}{\eqref{eq:15},\eqref{eq:16},\eqref{eq:18},\eqref{eq:27}}\\ \cline{2-3} \cline{5-5}
			& & & & &\\[-6pt]
			& \usym{2718}&\CheckmarkBold& &\usym{2718}&\\
			\cline{1-5}
			& & & & &\\[-6pt]
			\multirow{4}{*}{ES}&\usym{2718}&\usym{2718}&\multirow{2}{*}{Third line of \eqref{eq:8}}&\usym{2718}&\\ \cline{2-3} \cline{5-5}
			& & & & &\\[-6pt]
			&\usym{2718}&\CheckmarkBold& &\usym{2718}&\\ \cline{2-5}
			& & & & &\\[-6pt]
			&\CheckmarkBold&\usym{2718}&\multirow{2}{*}{\eqref{eq:10}}&\usym{2718}&\\ \cline{2-3} \cline{5-5}
			& & & & &\\[-6pt]
			&\CheckmarkBold&\CheckmarkBold& &\usym{2718}&\\
			\hline 
		\end{tabular}
	\end{threeparttable}
\end{table*}

Since individual variable in P2 is updated with the optimal closed-form solution under the BCD framework, the solution of P2 is a stationary point~\cite{Yang:20,Xuyang:13}. Together with the global optimal solution of P1 and \textbf{Proposition 1}, the solution generated by the \textbf{Algorithm 1} is at least a stationary point of the original problem. 

The complexity of updating $\bm x$ and $\bm \rho$ with closed-form expressions \eqref{eq:15}, \eqref{eq:16} is ${\rm{{\cal O}}}\left( {{{\left( {R + T} \right)} \mathord{\left/
			{\vphantom {{\left( {R + T} \right)} P}} \right.
			\kern-\nulldelimiterspace} P}} \right)$, where $P$ is the parallel processing factor since all these variables can be updated in parallel. For updating $\bm w$, according to \textbf{Lemma 4}, its complexity mainly come from the eigenvalue decomposition of $\bm \Xi$ and the bisection method of finding $\mu $, Therefore, its complexity is ${\rm{{\cal O}}}\left( {{N^3}} \right) + {\rm{{\cal O}}}\left( {\log \left( {\sqrt {{{{\rm Tr}\left( {\bm B} \right)} \mathord{\left/
					{\vphantom {{Tr\left( B \right)} {{\varepsilon ^2}{P_{BS}}}}} \right.
					\kern-\nulldelimiterspace} {{\varepsilon ^2}{P_{BS}}}}} } \right)} \right)$, where $\varepsilon $ is the accuracy of bisection search. For the update of RIS coefficient ${{\bm v}^{\mathcal t}}$ and ${{\bm v}^{\mathcal r}}$, since we sequentially update $M$ pairs of $\left\{ {v_m^{\mathcal t},v_m^{\mathcal r}} \right\}_{m = 1}^M$, and each pair involves a bisection search with range from 0 to ${{\pi  \mathord{\left/
				{\vphantom {\pi  2}} \right.
			\kern-\nulldelimiterspace} 2}}$, the corresponding complexity is ${\rm{{\cal O}}}\left( {M\log \left( {{\pi  \mathord{\left/
				{\vphantom {\pi  {2\varsigma }}} \right.
				\kern-\nulldelimiterspace} {2\varsigma }}} \right)} \right)$, where $\varsigma $ is the accuracy of bisection search. For the auxiliary variables ${{\bm \varphi} ^{\mathcal t}}$ and ${{\bm \varphi} ^{\mathcal r}}$, it can be parallelized and updated with closed-form according to \textbf{Lemmas 1} and \textbf{2}, with the complexity of this update being ${\rm{{\cal O}}}\left( {{M \mathord{\left/
			{\vphantom {M P}} \right.
			\kern-\nulldelimiterspace} P}} \right)$.

In total, the complexity of the penalty based BCD algorithm for this sum-rate maximization is ${I_{Pen}}{I_{BCD}}\left( {{\rm{{\cal O}}}\left( {{{\left( {R + T + M} \right)} \mathord{\left/
				{\vphantom {{\left( {R + T + M} \right)} P}} \right.
				\kern-\nulldelimiterspace} P} + {N^3}} \right) + {\rm{{\cal O}}}\left( {M\log \left( {{\pi  \mathord{\left/
					{\vphantom {\pi  {2\varsigma }}} \right.
					\kern-\nulldelimiterspace} {2\varsigma }}} \right)} \right)} \right.$ $+\left. {{\rm{{\cal O}}}\left( {\log \left( {\sqrt {{{{\rm Tr}\left( {\bm B} \right)} \mathord{\left/
						{\vphantom {{Tr\left( B \right)} {{\varepsilon ^2}{P_{BS}}}}} \right.
						\kern-\nulldelimiterspace} {{\varepsilon ^2}{P_{BS}}}}} } \right)} \right)} \right)$, where $I_{Pen}$ and $I_{BCD}$ are the number of penalty iterations and BCD iterations, respectively.

\section{Simulation Results and Discussions}
In this section, we evaluate the downlink sum-rate transmission performance through simulation. All problem instances are simulated using Matlab-R2023a on a Windows x64 desktop with 2.8 GHz CPU and 16 GB RAM, and the simulation results are obtained via averaging over 100 simulation trials. In the simulations, the BS and STAR-RIS are located at the coordinate (0,20m) and (40m,0), respectively. The STAR-RIS is placed along the y-axis and perpendicular to the ground. The reflection and transmission users are uniformly located within 8m of two sides of the STAR-RIS. The parameters $\left\{ {{\bm G},{{\bm h}_l},{{\bm d}_l}} \right\}$ are modeled as the Rician fading channels, which contain both the line-of-sight (LoS) and non-LoS (NLoS) components~\cite{Tse:05}. Take ${{\bm h}_l}$ as an example, the Rician fading channel model is ${{\bm h}_l} = \sqrt {{{{\nu _{{{\bm h}_l}}}} \mathord{\left/
			{\vphantom {{{\nu _{{{\bm h}_l}}}} {\left( {{\kappa _{\bm h}} + 1} \right)}}} \right.
			\kern-\nulldelimiterspace} {\left( {{\kappa _{\bm h}} + 1} \right)}}} \left( {\sqrt {{\kappa _{\bm h}}} {\bm h}_l^{LoS} + {\bm h}_l^{NLoS}} \right)$ and each parameter is specified below.
		
		\begin{enumerate}[leftmargin=0.5cm]
			\item ${\nu _{{{\bm h}_l}}} = {L_0}{\left( {{{{{\mathcal d}_{{{\bm h}_l}}}} \mathord{\left/
							{\vphantom {{{{\mathcal d}_{{{\bm h}_l}}}} {{{\mathcal d}_0}}}} \right.
							\kern-\nulldelimiterspace} {{{\mathcal d}_0}}}} \right)^{ - {\alpha _{\bm h}}}}$ is the distance dependent path-loss from RIS to the $l^{th}$ user, where ${L_0} =  - 30$dB denotes the path loss at the reference distance ${{\mathcal d}_0} = 1$m. ${{{\mathcal d}_{{{\bm h}_l}}}}$ is the distance between STAR-RIS and the $l^{th}$ user. ${{\alpha _{\bm h}}} = 2.2$ denotes the path-loss exponent for the RIS-user link. Correspondingly, the path-loss exponents for the BS-RIS link and the BS-user link are 2.2 and 3.6, respectively.
			\item ${{\kappa _{\bm h}}}$ is the Rician factor for the RIS-user link. A higher value of Rician factor means stronger LoS component. When the Rician factor is 0, it means there is no LoS signal and the channel reduces to the Rayleigh fading. In the simulations, the Rician factors for the RIS-user link ${{\kappa _{\bm h}}}$ is set to 5. Correspondingly, the Rician factor of the BS-RIS link and the BS-user link are 5 and 0, respectively.
			\item The LoS component ${\bm h}_l^{LoS}$ is modeled as the steering vector of the array responses. Hence the $m^{th}$ element ${{{\left[ {{\bm h}_l^{LoS}} \right]}_m}} = {e^{j{{2\pi \left( {m - 1} \right){d_A}\sin \left( \omega  \right)} \mathord{\left/
							{\vphantom {{2\pi \left( {m - 1} \right){d_A}\sin \left( \omega  \right)} \lambda }} \right.
							\kern-\nulldelimiterspace} \lambda }}}$, where $\omega $ denotes the angle-of-arrival (AoA) or
						angle-of-departure (AoD) of the array. In the simulation, ${{{d_A}} \mathord{\left/
						{\vphantom {{{d_A}} \lambda }} \right.
					\kern-\nulldelimiterspace} \lambda }=1/2$ and the $\omega $ is modeled as uniform distributed in $\left[ {0,2\pi } \right)$. On the other hand, ${{\bm h}_l^{NLoS}}$ denotes the NLoS component signal with each element obeying the normalized
				complex Gaussian distribution. 
		\end{enumerate}
	 To avoid repeating figure descriptions, the settings for ($M$, $N$, $K^{\mathcal r}$, $K^{\mathcal t}$, $L$, $P_{BS}$, $\sigma _l^2$) are provided in the caption of each figure.

\begin{figure}[t]
	\centering
	\includegraphics[width=1\linewidth]{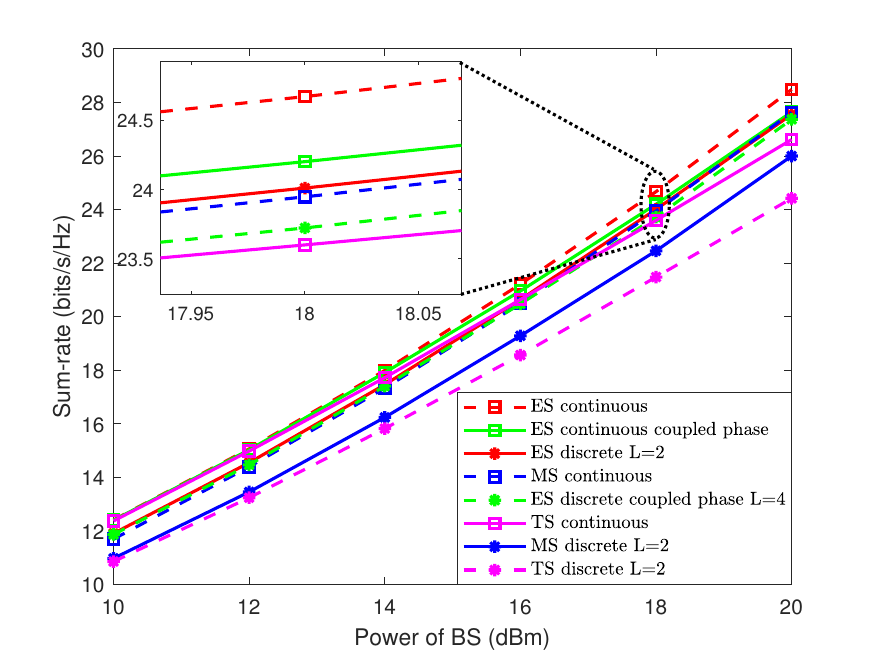}
	\label{fig:graph}
	\caption{Sum-rate comparison of the eight types of STAR-RIS at $M=30$, $N=16$, $K^{\mathcal r}=K^{\mathcal t}=4$ and $\sigma _l^2 =  - 80{\rm dBm}$}
\end{figure}

\begin{figure}[t]
	\centering
	\begin{minipage}[t]{0.5\textwidth}
		\centering
		\includegraphics[width=8.5cm]{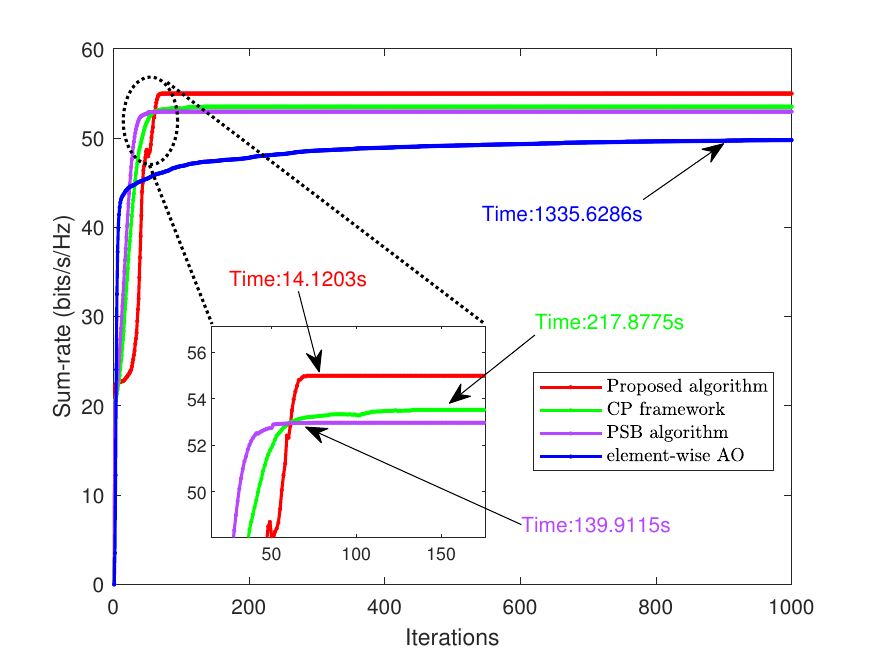}
		\centerline{\footnotesize {(a) Coupled-phase STAR-RIS}}
	\end{minipage}
	\begin{minipage}[t]{0.5\textwidth}
		\centering
		\includegraphics[width=8.5cm]{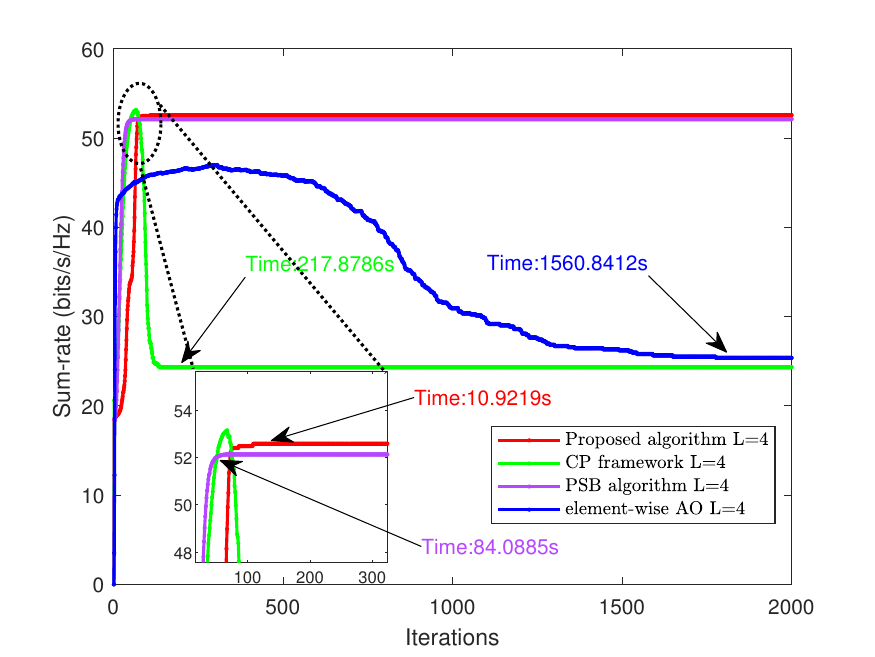}
		\centerline{\footnotesize {(b) Discrete coupled-phase STAR-RIS}}
	\end{minipage}
	\caption{Convergence behaviour with $M=30$, $N=16$, $K^{\mathcal r}=K^{\mathcal t}=4$, $P_{BS}=20{\rm dBm}$ and $\sigma _l^2 =  - 100{\rm dBm}$}
\end{figure}

Fig. 4 compares the sum-rate performance among the eight types of STAR-RIS using the proposed general optimization framework. Firstly, we can see that additional discrete phase constraint only slightly reduce the network throughput (2.94\% for two discrete phases in ES, 6.16\% and 8.31\% for two discrete phases in MS and TS, respectively), which is different from the finding in~\cite{Wu:22} that sparse phase (lower than 8 phases) will significantly affect the performance. The key reason is that the conclusion in~\cite{Wu:22} is based on quantization from continuous phase solution to its nearest discrete phase. Under the very few allowable discrete phases (e.g., in the simulation only 2 phases are allowed), the continuous phase and its closest discrete value are very different. Since the quantization is applied independently in each RIS element, the cumulated performance degradation will be significant compared to the continuous optimal solution. In contrast, the auxiliary variables ${{\bm \varphi} ^{\mathcal t}}$, ${{\bm \varphi} ^{\mathcal r}}$ in the proposed algorithm are only affected by the phase constraints, which can be regarded as the phase corrector for ${\bm v ^{\mathcal t}}$, ${\bm v^{\mathcal r}}$. Instead of directly quantizing the ${\bm v ^{\mathcal t}}$, ${\bm v^{\mathcal r}}$ to ${{\bm \varphi} ^{\mathcal t}}$, ${{\bm \varphi} ^{\mathcal r}}$, a penalty term is introduced to enforce an indirect quantization and it allows more freedom for $\bm v$ and $\bm \varphi$ to search for better solution. This avoids the significant performance loss and shows that quantization is a more influential factor than discrete phase that leads to performance degradation. Secondly, it is notice that the discrete phase constraint in MS and TS modes lead to more performance loss than that in the ES mode. This is probably because of the additional amplitude constraint of \eqref{eq:4c} imposed on MS and TS STAR-RIS coefficients. Without the amplitude constraint, ES STAR-RIS can optimize the amplitude to compensate the loss brought by the discrete phase constraint. Thirdly, from the magnified part of Fig. 4, we notice that the continuous coupled-phase constraint in STAR-RIS almost have no influence (only 0.89\% loss) on the system throughput compared to noncoupled-phase case, which is consistent with the conclusion in~\cite{Wang:23}. In contrast, the throughput degradation introduced by the 0-1 amplitude constraint in the MS mode is more obvious. This shows that amplitude constraint affects system performance more than the phase constraint. This insight is revealed for the first time due to the ease of comparison using the proposed unified penalty framework.

Figs. 5(a) and 5(b) focus on the ES STAR-RIS with coupled-phase (i.e., cases 7 and 8 in Table I) and compare the convergence behavior of the proposed algorithm with the elementwise-AO~\cite{Liu:22b}, the coupled-phase STAR-RIS framework (named as CP framework)~\cite{Wang:23} and penalty-based secrecy beamforming (PSB) algorithm~\cite{Zhang:23}. Although the original PSB algorithm was derived for secrecy sum-rate, we can modify it to handle sum-rate maximation. For the continuous coupled-phase case in Fig. 5(a), the proposed algorithm converges to the highest sum-rate with the shortest execution time. PSB algorithm and CP framework are slightly worse and the elementwise-AO performs the worst. As explained before \textbf{Proposition 1} (Section IIB), the penalty weight required in the CP and PSB algorithms would be larger than the proposed algorithm and slows down their convergence speeds. On the other hand, elementwise-AO needs to solve nonconvex subproblem on STAR-RIS element-wise level, which are handled by SCA and CVX, and thus incurs heavy computations. This is evidenced by the execution time at convergence marked in Fig. 5(a), where the proposed algorithm reduces almost a factor of 10 in computation time compared to the CP framework and the PSB algorithm, and almost a factor of 100 compared to the elementwise AO algorithm. Notice that since the CP framework, PSB algorithm and the proposed algorithm introduce penalty terms, the sum-rate may not be monotonic with respect to iterations as the monotonic property only holds for objective function including the penalty term.

For the discrete coupled-phase case in Fig. 5(b), the CP framework and elementwise-AO have noticeable performance decline when the number of iteration increases. This is due to the quantization to the continuous phase solution. This phenomenon coincides with the general statement from~\cite{Wu:22} that quantization in sparse phase (lower than 3 or 4 information bits) will strongly affect the performance. The PSB algorithm performs better than the elementwise-AO and CP framework because it employs quantization at RIS coefficient subproblem level. Compared with only one time quantization on the final continuous phase solution, repeated quantization at the level of RIS-subproblems avoids a sharp performance loss. Different from the above three methods, the proposed algorithm finds the global optimal closed-form solution at the RIS coefficient level subproblem even under the discrete phase constraint, which leads to the best throughput performance for the discrete and coupled-phase STAR-RIS in Fig. 5(b).

\begin{figure}[t]
	\centering
	\includegraphics[width=1\linewidth]{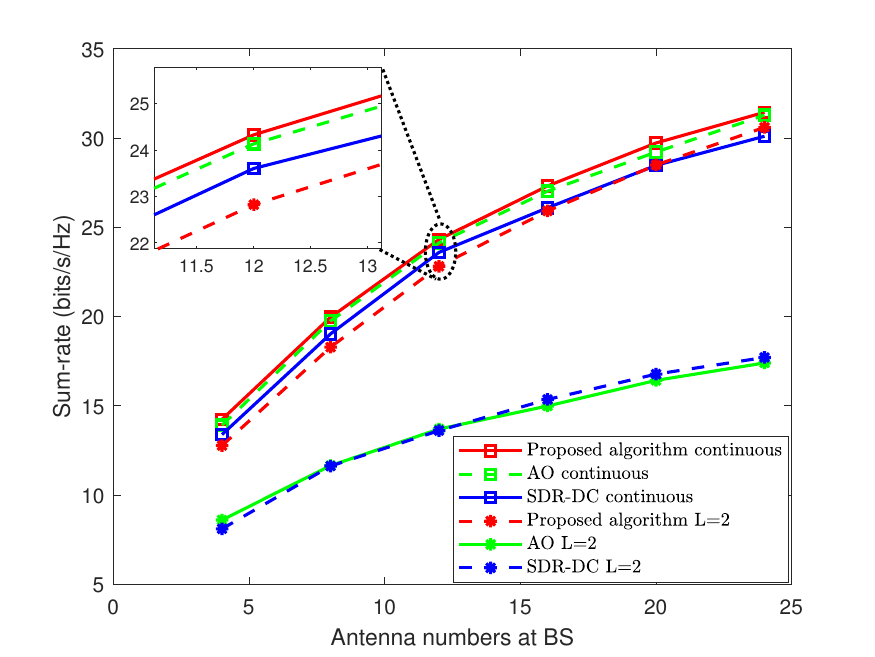}
	\label{fig:graph}
	\caption{Sum-rate of ES STAR-RIS with $M=30$, $K^{\mathcal r}=K^{\mathcal t}=4$, $P_{BS}=20{\rm dBm}$ and $\sigma _l^2 =  - 80{\rm dBm}$}
\end{figure}

Fig. 6 compares the sum-rate performance between the proposed algorithm, the AO algorithm~\cite{Perera:22} and SDR-DC method~\cite{Mu:22} for the basic ES STAR-RIS (i.e., the fifth and sixth cases in Table I).  Under continuous phase shift, the proposed algorithm performs the best and SDR-DC performs the worst, but the performance gap is not significant. However, under 2 discrete phases, the proposed algorithm performs only slight worse than continuous phase case (only 3.81\% degradation), and significantly outperforms the other two algorithms. This again verifies that discrete phase is not the major reason for performance degradation. It is just that the widely used quantization is not a good strategy for ES STAR-RIS under small number of phases.

\begin{figure}[t]
	\centering
	\includegraphics[width=1\linewidth]{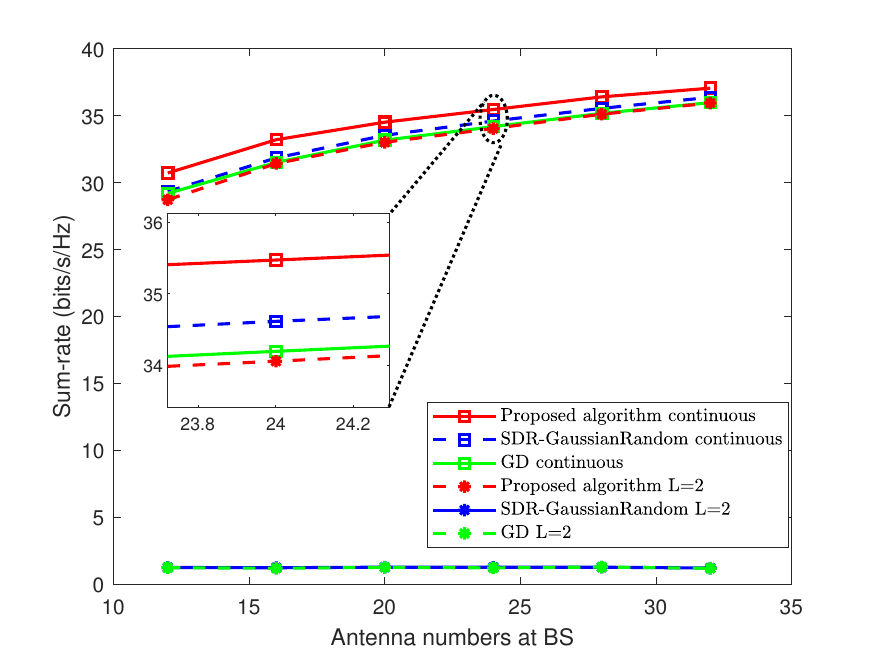}
	\label{fig:graph}
	\caption{Sum-rate of TS STAR-RIS with $M=40$, $K^{\mathcal r}=K^{\mathcal t}=5$, $P_{BS}=20{\rm dBm}$ and $\sigma _l^2 =  - 90{\rm dBm}$}
\end{figure}

Fig. 7 compares the sum-rate performance in TS STAR-RIS system (i.e., first two cases in Table I) among the proposed algorithm, the SDR algorithm~\cite{Mu:22} with Gaussian random, and gradient descent (GD) method~\cite{Du:23,Liu:23}. From Fig. 7, we can see that under continuous phase shift, the proposed algorithm performs the best, and then followed closely by SDR algorithm and GD method. However, under 2 discrete phase case, the GD method and SDR algorithm fail completely.

\begin{figure}[t]
	\centering
	\includegraphics[width=1\linewidth]{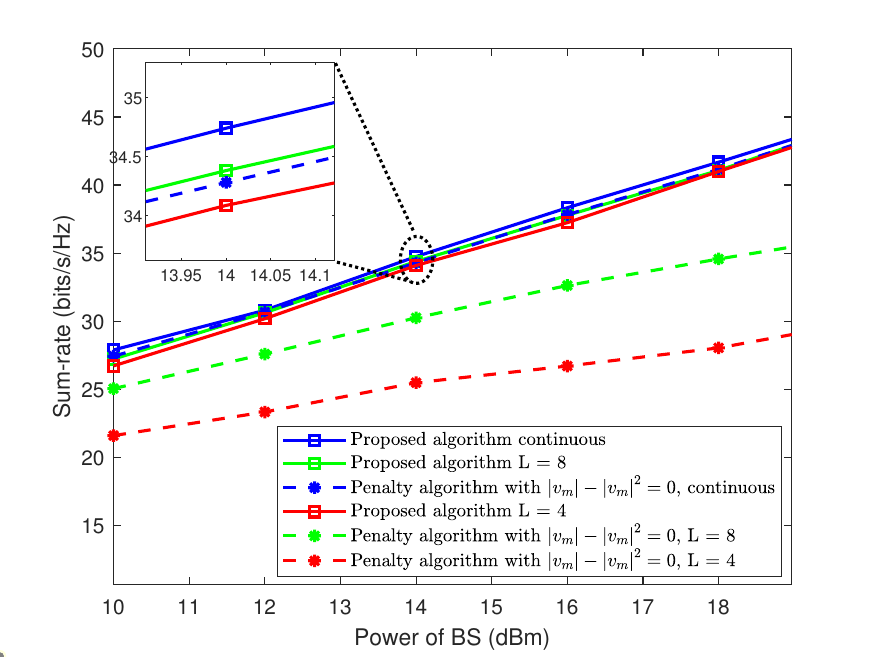}
	\label{fig:graph}
	\caption{Sum-rate of MS STAR-RIS with $M=20$, $K^{\mathcal r}=K^{\mathcal t}=5$,  and $\sigma _l^2 =  - 80{\rm dBm}$}
\end{figure}

Fig. 8 compares the sum-rate performance between the proposed algorithm and the direct penalty algorithm with the widely used penalty term $\left| {v_m^{\mathcal k}} \right| - {\left| {v_m^{\mathcal k}} \right|^2} = 0,\forall {\mathcal k} \in \left\{ {{\mathcal t},{\mathcal r}} \right\},m \in \left\{ {1, \cdots ,M} \right\}$~\cite{Mu:22,Zhang:23} for MS STAR-RIS (i.e., the third and fourth cases in Table I). Since no existing literature studies the MS STAR-RIS under discrete phase, we add a quantization step after the direct penalty algorithm to make the discrete phase constraint satisfied. From Fig. 8, we observe that when there is no discrete phase constraint, the proposed auxiliary variable based penalty method outperforms the direct penalty method, and the performance of direct penalty method is even worse than the proposed algorithm with discrete phase $L=8$. This shows the possibility of exploring better penalty term to improve the performance of continuous phase MS STAR-RIS. In additional, the discrete phase constraint just slightly degrades the performance (2.28\% for $L=4$ and 1.27\% for $L=8$) if the proposed algorithm is employed, which coincides with the conclusion in ES STAR-RIS that the discrete phase is not the major reason for performance degradation. In contrast, the quantization adopted in discrete MS STAR-RIS heavily impairs the performance (28.3\% degradation at $L=4$ and 14.1\% degradation at $L=8$). Besides, quantization in MS STAR-RIS makes the sum-rate grows slowly with the power of BS. This all shows that quantization is not a good option for MS STAR-RIS under discrete phase.

\begin{figure}[t]
	\centering
	\includegraphics[width=1\linewidth]{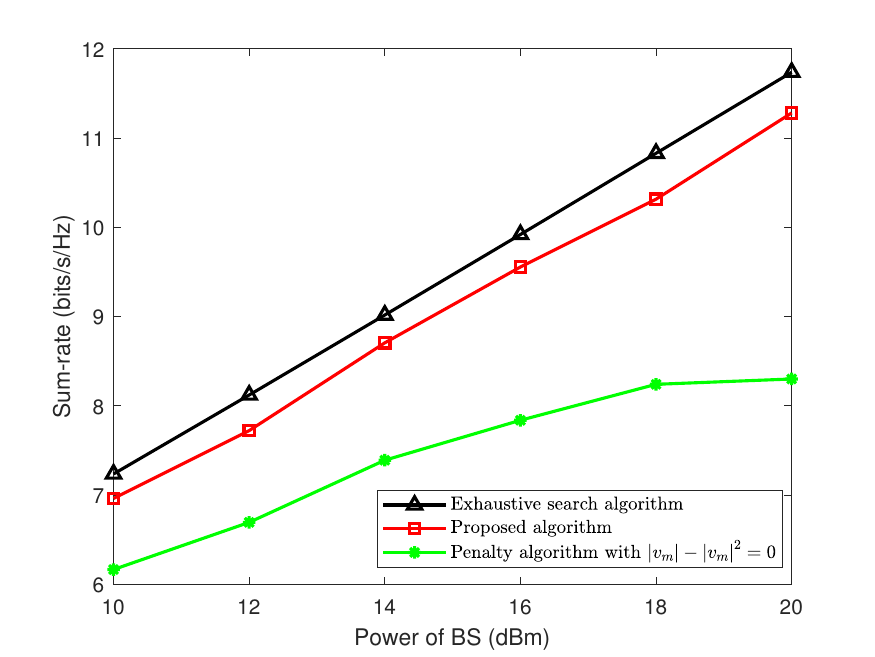}
	\label{fig:graph}
	\caption{Sum-rate of MS STAR-RIS with $L=2$, $M=6$, $N=4$, $K^{\mathcal r}=K^{\mathcal t}=1$ and $\sigma _l^2 =  - 80{\rm dBm}$}
\end{figure}
Finally, we evaluate the performance of the proposed algorithm when compared to exhaustive search solution. In particular, we consider the discrete phase MS STAR-RIS case, in which both amplitude and phase are discrete variables. For a MS STAR-RIS with $M$ elements and $L$ discrete phase, it will have ${\left( {2L} \right)^M}$ different combinations, which is exponential with $M$. Therefore, exhaustive search algorithm is only possible for small $M$ and $L$. Fig. 9 shows the results of $L=2$ and $M=6$. It can be seen that the performance degradation of the proposed method from the optimal solution using exhaustive searching is only 2.4\%$\sim $4.3\%, and their sum-rates grow at the same rate as the power of BS increases. However, for the direct penalty algorithm with $\left| {v_m^{\mathcal k}} \right| - {\left| {v_m^{\mathcal k}} \right|^2} = 0,\forall {\mathcal k} \in \left\{ {{\mathcal t},{\mathcal r}} \right\}$, the degradation is more significant (14.7\%$\sim $29.2\%) and shows slower rate of sum-rate increase with power of BS.

\section{Conclusions}
This paper proposed a unified framework to efficiently handle the constraints introduced by various kinds of STAR-RISs, even with discrete and operating mode constraints. The proposed unified framework consists of introducing auxiliary variables for the STAR-RIS phases such that closed-form global optimal solution is possible at the subproblem level. In addition to unifying the algorithm derivations for systems involving various kinds of STAR-RISs and lowering the computational complexity, convergence of such framework are established theoretically. As an illustrated example, a downlink STAR-RIS assisted transmission system was investigated under the proposed framework. Simulation results showed that the proposed framework outperforms other existing state-of-the-art methods, and revealed for the first time that the existence of discrete phase may not cause significant performance degradation.

\bibliographystyle{IEEEtran}

\bibliography{MISOtransmissionSystem2}

\begin{thebibliography}{10}
\providecommand{\url}[1]{#1}
\csname url@samestyle\endcsname
\providecommand{\newblock}{\relax}
\providecommand{\bibinfo}[2]{#2}
\providecommand{\BIBentrySTDinterwordspacing}{\spaceskip=0pt\relax}
\providecommand{\BIBentryALTinterwordstretchfactor}{4}
\providecommand{\BIBentryALTinterwordspacing}{\spaceskip=\fontdimen2\font plus
\BIBentryALTinterwordstretchfactor\fontdimen3\font minus
  \fontdimen4\font\relax}
\providecommand{\BIBforeignlanguage}[2]{{%
\expandafter\ifx\csname l@#1\endcsname\relax
\typeout{** WARNING: IEEEtran.bst: No hyphenation pattern has been}%
\typeout{** loaded for the language `#1'. Using the pattern for}%
\typeout{** the default language instead.}%
\else
\language=\csname l@#1\endcsname
\fi
#2}}
\providecommand{\BIBdecl}{\relax}
\BIBdecl

\bibitem{Wu:19}
Q.~Wu and R.~Zhang, ``Intelligent reflecting surface enhanced wireless network
  via joint active and passive beamforming,'' \emph{IEEE Trans. Wirel.
  Commun.}, vol.~18, no.~11, pp. 5394--5409, Nov. 2019.

\bibitem{Kudathanthirige:20}
D.~Kudathanthirige, D.~Gunasinghe, and G.~Amarasuriya, ``Performance analysis
  of intelligent reflective surfaces for wireless communication,'' in \emph{ICC
  2020}, July. 2020, pp. 1--6.

\bibitem{Basar:19}
E.~Basar, M.~Di~Renzo, J.~De~Rosny, M.~Debbah, M.-S. Alouini, and R.~Zhang,
  ``Wireless communications through reconfigurable intelligent surfaces,''
  \emph{IEEE Access}, vol.~7, pp. 116\,753--116\,773, Aug. 2019.

\bibitem{Chen:19}
J.~Chen, Y.-C. Liang, Y.~Pei, and H.~Guo, ``Intelligent reflecting surface: A
  programmable wireless environment for physical layer security,'' \emph{IEEE
  Access}, vol.~7, pp. 82\,599--82\,612, June. 2019.

\bibitem{Renzo:19}
M.~Renzo, M.~Debbah, D.~Phan-Huy, and et~al, ``Smart radio enviroments
  empowered by reconfigurable ai meta-surfaces: An idea whose time has come,''
  \emph{J Wireless Com Network}, no. 129, pp. 129--149, May. 2019.

\bibitem{Li:22}
Z.~Li, S.~Wang, M.~Wen, and Y.-C. Wu, ``Secure multicast energy-efficiency
  maximization with massive riss and uncertain csi: First-order algorithms and
  convergence analysis,'' \emph{IEEE Trans. Wirel. Commun.}, vol.~21, no.~9,
  pp. 6818--6833, Sept. 2022.

\bibitem{Yang:22}
Y.~Yang, Y.~Gong, and Y.-C. Wu, ``Intelligent-reflecting-surface-aided mobile
  edge computing with binary offloading: Energy minimization for iot devices,''
  \emph{IEEE Internet Things J.}, vol.~9, no.~15, pp. 12\,973--12\,983, May.
  2022.

\bibitem{Xu:21}
J.~Xu, Y.~Liu, X.~Mu, and O.~A. Dobre, ``Star-riss: Simultaneous transmitting
  and reflecting reconfigurable intelligent surfaces,'' \emph{IEEE Commun.
  Lett.}, vol.~25, no.~9, pp. 3134--3138, May. 2021.

\bibitem{Wu:21}
C.~Wu, Y.~Liu, X.~Mu, X.~Gu, and O.~A. Dobre, ``Coverage characterization of
  star-ris networks: Noma and oma,'' \emph{IEEE Commun. Lett.}, vol.~25, no.~9,
  pp. 3036--3040, Sept. 2021.

\bibitem{Niu:21}
H.~Niu, Z.~Chu, F.~Zhou, and Z.~Zhu, ``Simultaneous transmission and reflection
  reconfigurable intelligent surface assisted secrecy miso networks,''
  \emph{IEEE Commun. Lett.}, vol.~25, no.~11, pp. 3498--3502, Aug. 2021.

\bibitem{Alom:17}
M.~Z. Alom, B.~Van~Essen, A.~T. Moody, D.~P. Widemann, and T.~M. Taha,
  ``Quadratic unconstrained binary optimization (qubo) on neuromorphic
  computing system,'' in \emph{IJCNN-2017}, July. 2017, pp. 3922--3929.

\bibitem{Perera:22}
P.~P. Perera, V.~G. Warnasooriya, D.~Kudathanthirige, and H.~A. Suraweera,
  ``Sum rate maximization in star-ris assisted full-duplex communication
  systems,'' in \emph{ICC 2022}, Aug. 2022, pp. 3281--3286.

\bibitem{Mu:22}
X.~Mu, Y.~Liu, L.~Guo, J.~Lin, and R.~Schober, ``Simultaneously transmitting
  and reflecting (star) ris aided wireless communications,'' \emph{IEEE Trans.
  Wirel. Commun.}, vol.~21, no.~5, pp. 3083--3098, May. 2022.

\bibitem{Liu:22a}
Y.~Liu, J.~Xu, and X.~Mu, ``Fast beam splitting technique for star-riss with
  coupled t\&r phase shifts,'' in \emph{AT-AP-RASC 2022}, May. 2022, pp. 1--3.

\bibitem{Zhong:22}
R.~Zhong, Y.~Liu, X.~Mu, Y.~Chen, X.~Wang, and L.~Hanzo, ``Hybrid reinforcement
  learning for star-riss: A coupled phase-shift model based beamformer,''
  \emph{IEEE J. Sel. Areas Commun.}, vol.~40, no.~9, pp. 2556--2569, Sept.
  2022.

\bibitem{Liu:22b}
Y.~Liu, X.~Mu, R.~Schober, and H.~V. Poor, ``Simultaneously transmitting and
  reflecting (star)-riss: A coupled phase-shift model,'' in \emph{ICC 2022},
  May. 2022, pp. 2840--2845.

\bibitem{Wang:23}
Z.~Wang, X.~Mu, Y.~Liu, and R.~Schober, ``Coupled phase-shift star-riss: A
  general optimization framework,'' \emph{IEEE Wireless Commun. Lett.},
  vol.~12, no.~2, pp. 207--211, Nov. 2023.

\bibitem{Chen:21}
Y.~Chen, B.~Ai, H.~Zhang, Y.~Niu, L.~Song, Z.~Han, and H.~Vincent~Poor,
  ``Reconfigurable intelligent surface assisted device-to-device
  communications,'' \emph{IEEE Trans. Wirel. Commun.}, vol.~20, no.~5, pp.
  2792--2804, May. 2021.

\bibitem{Gao:21}
H.~Gao, K.~Cui, C.~Huang, and C.~Yuen, ``Robust beamforming for ris-assisted
  wireless communications with discrete phase shifts,'' \emph{IEEE Wireless
  Commun. Lett.}, vol.~10, no.~12, pp. 2619--2623, Aug. 2021.

\bibitem{Katwe:23}
M.~Katwe, K.~Singh, B.~Clerckx, and C.-P. Li, ``Improved spectral efficiency in
  star-ris aided uplink communication using rate splitting multiple access,''
  \emph{IEEE Trans. Wirel. Commun.}, pp. 1--1, Jan. 2023.

\bibitem{Qin:23}
X.~Qin, Z.~Song, T.~Hou, W.~Yu, J.~Wang, and X.~Sun, ``Joint resource
  allocation and configuration design for star-ris-enhanced wireless-powered
  mec,'' \emph{IEEE Trans Commun}, vol.~71, no.~4, pp. 2381--2395, Jan. 2023.

\bibitem{Du:23}
W.~Du, Z.~Chu, G.~Chen, P.~Xiao, Y.~Xiao, X.~Wu, and W.~Hao, ``Star-ris
  assisted wireless powered iot networks,'' \emph{IEEE Trans. Veh. Technol.},
  pp. 1--15, Mar. 2023.

\bibitem{Wuyu:22}
C.~Wu, C.~You, Y.~Liu, X.~Gu, and Y.~Cai, ``Channel estimation for
  star-ris-aided wireless communication,'' \emph{IEEE Commun. Lett.}, vol.~26,
  no.~3, pp. 652--656, Dec. 2022.

\bibitem{Ni:22}
W.~Ni, Y.~Liu, Y.~C. Eldar, Z.~Yang, and H.~Tian, ``Star-ris integrated
  nonorthogonal multiple access and over-the-air federated learning: Framework,
  analysis, and optimization,'' \emph{IEEE Internet Things J.}, July. 2022.

\bibitem{Zhang:22b}
Q.~Zhang, Y.~Zhao, H.~Li, S.~Hou, and Z.~Song, ``Joint optimization of star-ris
  assisted uav communication systems,'' \emph{IEEE Wireless Commun. Lett.},
  vol.~11, no.~11, pp. 2390--2394, Sept. 2022.

\bibitem{Fang:23}
F.~Fang, B.~Wu, S.~Fu, Z.~Ding, and X.~Wang, ``Energy-efficient design of
  star-ris aided mimo-noma networks,'' \emph{IEEE Trans Commun}, vol.~71,
  no.~1, pp. 498--511, Nov. 2023.

\bibitem{Zhang:22}
Z.~Zhang, J.~Chen, Y.~Liu, Q.~Wu, B.~He, and L.~Yang, ``On the secrecy design
  of star-ris assisted uplink noma networks,'' \emph{IEEE Trans. Wirel.
  Commun.}, vol.~21, no.~12, pp. 11\,207--11\,221, Dec. 2022.

\bibitem{Abrar:23}
M.~F.~U. Abrar, M.~Talha, R.~I. Ansari, S.~A. Hassan, and H.~Jung,
  ``Optimization of star-ris-assisted hybrid noma mmwave communication,''
  \emph{IEEE Trans. Veh. Technol.}, pp. 1--16, Mar. 2023.

\bibitem{Zhao:22}
J.~Zhao, Y.~Zhu, X.~Mu, K.~Cai, Y.~Liu, and L.~Hanzo, ``Simultaneously
  transmitting and reflecting reconfigurable intelligent surface (star-ris)
  assisted uav communications,'' \emph{IEEE J. Sel. Areas Commun.}, vol.~40,
  no.~10, pp. 3041--3056, Aug. 2022.

\bibitem{Zhai:23}
X.~Zhai, G.~Han, Y.~Cai, Y.~Liu, and L.~Hanzo, ``Simultaneously transmitting
  and reflecting (star) ris assisted over-the-air computation systems,''
  \emph{IEEE Trans Commun}, vol.~71, no.~3, pp. 1309--1322, Mar. 2023.

\bibitem{Zhang:23}
Z.~Zhang, Z.~Wang, Y.~Liu, B.~He, L.~Lv, and J.~Chen, ``Security enhancement
  for coupled phase-shift star-ris networks,'' \emph{IEEE Trans. Veh.
  Technol.}, vol.~72, no.~6, pp. 8210--8215, Feb. 2023.

\bibitem{Bedi:22}
A.~S. Bedi, K.~Rajawat, V.~Aggarwal, and A.~Koppel, ``Escaping saddle points
  for successive convex approximation,'' \emph{IEEE Trans. Signal Process.},
  vol.~70, pp. 307--321, Dec. 2022.

\bibitem{Li:19}
Y.~Li, M.~Xia, and Y.-C. Wu, ``Energy-efficient precoding for non-orthogonal
  multicast and unicast transmission via first-order algorithm,'' \emph{IEEE
  Trans. Wirel. Commun.}, vol.~18, no.~9, pp. 4590--4604, July. 2019.

\bibitem{beck:17}
\BIBentryALTinterwordspacing
A.~Beck, \emph{First-Order methods in optimization}.\hskip 1em plus 0.5em minus
  0.4em\relax Philadelphia, PA: Society for Industrial and Applied Mathematics,
  2017. [Online]. Available:
  \url{https://epubs.siam.org/doi/abs/10.1137/1.9781611974997}
\BIBentrySTDinterwordspacing

\bibitem{Zong:21}
Z.~Li, M.~Xia, M.~Wen, and Y.-C. Wu, ``Massive access in secure noma under
  imperfect csi: Security guaranteed sum-rate maximization with first-order
  algorithm,'' \emph{IEEE J. Sel. Areas Commun.}, vol.~39, no.~4, pp.
  998--1014, Apr. 2021.

\bibitem{Hou:22}
T.~Hou, J.~Wang, Y.~Liu, X.~Sun, A.~Li, and B.~Ai, ``A joint design for
  star-ris enhanced noma-comp networks: A
  simultaneous-signal-enhancement-and-cancellation-based (ssecb) design,''
  \emph{IEEE Trans. Veh. Technol.}, vol.~71, no.~1, pp. 1043--1048, Jan. 2022.

\bibitem{Lee:15}
H.~Lee, K.-J. Lee, H.~Kim, B.~Clerckx, and I.~Lee, ``Resource allocation
  techniques for wireless powered communication betworks with energy storage
  constraint,'' \emph{IEEE Trans. Wirel. Commun.}, vol.~15, no.~4, pp.
  2619--2628, Apr. 2016.

\bibitem{Ju:14}
H.~Ju and R.~Zhang, ``Optimal resource allocation in full-duplex
  wireless-powered communication network,'' \emph{IEEE Transactions on
  Communications}, vol.~62, no.~10, pp. 3528--3540, Oct. 2014.

\bibitem{Shen:18}
K.~Shen and W.~Yu, ``Fractional programming for communication systems—part
  ii: Uplink scheduling via matching,'' \emph{IEEE Trans. Signal Process.},
  vol.~66, no.~10, pp. 2631--2644, Mar. 2018.

\bibitem{Yang:20}
Y.~Yang, M.~Pesavento, Z.-Q. Luo, and B.~Ottersten, ``Inexact block coordinate
  descent algorithms for nonsmooth nonconvex optimization,'' \emph{IEEE Trans.
  Signal Process.}, vol.~68, pp. 947--961, Dec. 2020.

\bibitem{Xuyang:13}
Y.~Xu and W.~Yin, ``A block coordinate descent method for regularized
  multiconvex optimization with applications to nonnegative tensor
  factorization and completion,'' \emph{SIAM J. Imaging Sci.}, vol.~6, no.~3,
  pp. 1758--1789, 2013.

\bibitem{Tse:05}
D.~Tse and P.~Viswanath, \emph{Fundamentals of wireless communication}.\hskip
  1em plus 0.5em minus 0.4em\relax Cambridge university press, 2005.

\bibitem{Wu:22}
C.~Wu, X.~Mu, Y.~Liu, X.~Gu, and X.~Wang, ``Resource allocation in
  star-ris-aided networks: Oma and noma,'' \emph{IEEE Trans. Wirel. Commun.},
  vol.~21, no.~9, pp. 7653--7667, Mar. 2022.

\bibitem{Liu:23}
Z.~Liu, Z.~Li, M.~Wen, Y.~Gong, and Y.-C. Wu, ``Star-ris aided mobile edge
  computing: Computation rate maximization with binary amplitude
  coefficients,'' \emph{IEEE Trans Commun}, vol.~71, no.~7, pp. 4313--4327,
  July. 2023.

\bibitem{bartle:00}
R.~G. Bartle and D.~R. Sherbert, \emph{Introduction to real analysis}.\hskip
  1em plus 0.5em minus 0.4em\relax John Wiley \& Sons, Inc., 2000.

\bibitem{kiwiel:01}
K.~C. Kiwiel, ``Convergence and efficiency of subgradient methods for
  quasiconvex minimization,'' \emph{Math Program}, vol.~90, pp. 1--25, Mar.
  2001.

\bibitem{bolte:14}
J.~Bolte, S.~Sabach, and M.~Teboulle, ``Proximal alternating linearized
  minimization for nonconvex and nonsmooth problems,'' \emph{Math Program},
  vol. 146, no. 1-2, pp. 459--494, July. 2014.

\bibitem{Guo:20}
H.~Guo, Y.-C. Liang, J.~Chen, and E.~G. Larsson, ``Weighted sum-rate
  maximization for reconfigurable intelligent surface aided wireless
  networks,'' \emph{IEEE Trans. Wirel. Commun.}, vol.~19, no.~5, pp.
  3064--3076, May. 2020.

\end{thebibliography}

\clearpage
\begin{center}
	{{\normalsize S}{\small UPPLEMENTARY} {\normalsize M}{\small ATERIAL}}
\end{center}
\section{Appendix}

\subsection{Proof of Proposition 1}
Let ${\bm p}_\gamma ^n = \left[ {{\bm z}_\gamma ^n,\lambda_\gamma ^{{\mathcal t},n},\lambda_\gamma ^{{\mathcal r},n}} \right]$, ${\bm v}_\gamma ^n = \left[ {{\bm v}_\gamma ^{{\mathcal t},n},{\bm v}_\gamma ^{{\mathcal r},n}} \right]$ and ${\bm \varphi} _\gamma ^n = \left[ {{\bm \varphi} _\gamma ^{{\mathcal t},n},{\bm \varphi} _\gamma ^{{\mathcal r},n}} \right]$ as the solutions at the ${n^{th}}$ BCD iteration under the penalty $\gamma$. Using the above compact notations, the penalty term at the ${n^{th}}$ BCD iteration under the penalty $\gamma$ is reformulated as ${\rm{{\cal G}}}\left( {{{\bm v}_\gamma^n },{{\bm \varphi} _\gamma^n }} \right) = \frac{\gamma }{2}{{\left| {{{\bm v}^{n}_\gamma} - {{\bm \varphi} ^{n}_\gamma}} \right|}_2^2}$. Since the BCD iteration is monotonic under any fixed $\gamma$, we have 
\begin{equation}\label{eq:28}
	{\rm{{\cal G}}}\left( {{\bm v}_\gamma ^n,{\bm \varphi} _\gamma ^n} \right) \le {\rm{{\cal F}}}\left( {{\bm p}_\gamma ^0,{\bm v}_\gamma ^0} \right) + {\rm{{\cal G}}}\left( {{\bm v}_\gamma ^0,{\bm \varphi} _\gamma ^0} \right) - {\rm{{\cal F}}}\left( {{\bm p}_\gamma ^n,{\bm v}_\gamma ^n} \right).
\end{equation} 
With $\left\{ {{\bm p}_\gamma ^0,{\bm v}_\gamma ^0,{\bm \varphi} _\gamma ^0} \right\}$ being the initial point of the BCD iteration, we can always select them so that the first two terms of the right hand side of \eqref{eq:28} are bounded. Besides, ${\rm{{\cal F}}}$ is bounded from below in order to make its minimization meaningful. This makes $- {\rm{{\cal F}}}\left( {{\bm p}_\gamma ^n,{\bm v}_\gamma ^n} \right)$ bounded from above, so does the right hand side of \eqref{eq:28}. Therefore, the left hand side of \eqref{eq:28} is bounded from above for all $n$. That is, $\frac{\gamma }{2}{{\left| {{{\bm v}^{n}_\gamma} - {{\bm \varphi} ^{n}_\gamma}} \right|}_2^2}$ is bounded from above for all $n$. On the other hand, since the \textbf{Proposition 1} assumes that the $\left\{ {{\bm p}_\gamma ^n,{\bm v}_\gamma ^n} \right\}$ of P2 does not contains infinite value, there exists a sufficient large $D$ so that $\left| {\left[ {{\bm p}_\gamma ^n,{\bm v}_\gamma ^n} \right]} \right| \le D$ for all $n$ (otherwise, $\left| {\left[ {{\bm p}_\gamma ^n,{\bm v}_\gamma ^n} \right]} \right| \to \infty $ for some $n$ and it contains infinite point, which violates the assumption in\textbf{ Proposition 1}). Therefore, we obtain ${\bm \varphi}_\gamma ^n$ is bounded for all $n$. Notice that the solution ${{\bm p}_\gamma ^n,{\bm v}_\gamma ^n}$ are also bounded by $D$, there must exist a subsequence ${\left\{ {{\bm p}_\gamma ^{n_j},{\bm v}_\gamma ^{n_j},{\bm \varphi} _\gamma ^{n_j}} \right\}_{j \in {\rm{{\mathbb N}}}}}$ converges to some limit point ${\left\{ {{\bm p}_\gamma ^*,{\bm v}_\gamma ^*,{\bm \varphi} _\gamma ^*} \right\}}$ based on Bolzano-Weierstrass Theorem~\cite{bartle:00}.

For the ease of representation, the constraints of P2 are denoted using an indicator function ${\mathbbm{I}_1}\left( {{\bm p},{\bm v}} \right) = \left\{ \begin{array}{l}
	0,\;\;\;{\rm if\;the\;constraints\;in\;P2\;hold}\\
	\infty ,\;{\rm otherwise}
\end{array} \right.$. By assumption in \textbf{Proposition 1} that the solution ${{\bm p}^{n_j}_\gamma}$, ${\bm v}_\gamma ^{n_j}$ of P2 is a stationary point, the first-order optimality condition holds~\cite{beck:17}, which is shown in \eqref{eq:29},
\begin{figure*}
	\begin{equation}\label{eq:29}
		\left\langle {\left[ {\begin{array}{*{20}{c}}
					{{\nabla _{\bm p}}{\rm{{\cal F}}}\left( {{\bm p}_\gamma ^{n_j},{\bm v}_\gamma ^{n_j}} \right) + {\partial _{\bm p}}{\mathbbm{I}_1}\left( {{\bm p}_\gamma ^{n_j},{\bm v}_\gamma ^{n_j}} \right)}\\
					{{\nabla _{\bm v}}{\rm{{\cal F}}}\left( {{\bm p}_\gamma ^{n_j},{\bm v}_\gamma ^{n_j}} \right) + {\nabla _{\bm v}}{\rm{{\cal G}}}\left( {{\bm v}_\gamma ^{n_j},{\bm \varphi} _\gamma ^{{n_j} - 1}} \right) + {\partial _{\bm v}}{\mathbbm{I}_1}\left( {{\bm p}_\gamma ^{n_j},{\bm v}_\gamma ^{n_j}} \right)}
			\end{array}} \right],\left[ {\begin{array}{*{20}{c}}
					{{\bm p} - {\bm p}_\gamma ^{n_j}}\\
					{{\bm v} - {\bm v}_\gamma ^{n_j}}
			\end{array}} \right]} \right\rangle  \ge 0, \forall {\bm p},{\bm v},
	\end{equation}
\end{figure*}
where ${\partial _{\bm p}}{\mathbbm{I}_1}\left( {{\bm p},{\bm v}} \right)$ and ${\partial _{\bm v}}{\mathbbm{I}_1}\left( {{\bm p},{\bm v}} \right)$ are the limiting subdifferential of the non-smooth function ${\mathbbm{I}_1}\left( {{\bm p},{\bm v}} \right)$ with respect to $\bm p$ and $\bm v$, respectively.

Since ${\left\{ {{\bm p}_\gamma ^{{n_j}},{\bm v}_\gamma ^{{n_j}}} \right\}_{j \in {\rm{{\mathbb N}}}}}$ and ${\left\{ {{\bm p}_\gamma ^*,{\bm v}_\gamma ^*} \right\}}$ are feasible points of P2, we have $\mathop {\lim }\limits_{j \to \infty } {\partial _{\bm p}}{\mathbbm{I}_1}\left( {{\bm p}_\gamma ^{{n_j}},{\bm v}_\gamma ^{{n_j}}} \right) = {\partial _{\bm p}}{\mathbbm{I}_1}\left( {\mathop {\lim }\limits_{j \to \infty } {\bm p}_\gamma ^{{n_j}},{\bm v}_\gamma ^{{n_j}}} \right) = {\partial _{\bm p}}{\mathbbm{I}_1}\left( {{\bm p}_\gamma ^*,{\bm v}_\gamma ^*} \right)$ and $\mathop {\lim }\limits_{j \to \infty } {\partial _{\bm p}}{\mathbbm{I}_1}\left( {{\bm p}_\gamma ^{{n_j}},{\bm v}_\gamma ^{{n_j}}} \right) = {\partial _{\bm p}}{\mathbbm{I}_1}\left( {{\bm p}_\gamma ^*,{\bm v}_\gamma ^*} \right)$. Then, taking $j \to \infty $ in \eqref{eq:29}, we obtain \eqref{eq:30},
\begin{figure*}
	\begin{equation}\label{eq:30}
		\left\langle {\left[ {\begin{array}{*{20}{c}}
					{{\nabla _{\bm p}}{\rm{{\cal F}}}\left( {{\bm p}_\gamma ^*,{\bm v}_\gamma ^*} \right) + {\partial _{\bm p}}{\mathbbm{I}_1}\left( {{\bm p}_\gamma ^*,{\bm v}_\gamma ^*} \right)}\\
					{{\nabla _{\bm v}}{\rm{{\cal F}}}\left( {{\bm p}_\gamma ^*,{\bm v}_\gamma ^*} \right) + {\nabla _{\bm v}}{\rm{{\cal G}}}\left( {{\bm v}_\gamma ^*,{\bm \varphi} _\gamma ^*} \right) + {\partial _{\bm v}}{\mathbbm{I}_1}\left( {{\bm p}_\gamma ^*,{\bm v}_\gamma ^*} \right)}
			\end{array}} \right],\left[ {\begin{array}{*{20}{c}}
					{{\bm p} - {\bm p}_\gamma ^*}\\
					{{\bm v} - {\bm v}_\gamma ^*}
			\end{array}} \right]} \right\rangle  \ge 0,\;\; \forall {\bm p},{\bm v},
	\end{equation}
\end{figure*}
which can be written as
\begin{equation}\label{eq:31}
	{\bm 0} \in \left[ {\begin{array}{*{20}{c}}
			{{\nabla _{\bm p}}{\rm{{\cal F}}}\left( {{\bm p}_\gamma ^*,{\bm v}_\gamma ^*} \right) + {\partial _{\bm p}}{\mathbbm{I}_1}\left( {{\bm p}_\gamma ^*,{\bm v}_\gamma ^*} \right)}\\
			{{\nabla _{\bm v}}{\rm{{\cal F}}}\left( {{\bm p}_\gamma ^*,{\bm v}_\gamma ^*} \right) + {\nabla _{\bm v}}{\rm{{\cal G}}}\left( {{\bm v}_\gamma ^*,{\bm \varphi} _\gamma ^*} \right) + {\partial _{\bm v}}{\mathbbm{I}_1}\left( {{\bm p}_\gamma ^*,{\bm v}_\gamma ^*} \right)}
	\end{array}} \right].
\end{equation}

On the other hand, since by the assumption in \textbf{Proposition 1} that the global optimal solution of P1 is obtained, we have
\begin{equation}\label{eq:32}
	{\rm{{\cal G}}}\left( {{\bm v}_\gamma ^{n_j},{\bm \varphi} } \right) + {\mathbbm{I}_2}\left( {\bm \varphi}  \right) \ge {\rm{{\cal G}}}\left( {{\bm v}_\gamma ^{n_j},{\bm \varphi} _\gamma ^{n_j}} \right) + {\mathbbm{I}_2}\left( {{\bm \varphi} _\gamma ^{n_j}} \right),\;\;\forall {\bm \varphi},
\end{equation}
where ${\mathbbm{I}_2}\left( {\bm \varphi } \right) = \left\{ \begin{array}{l}
	0,\;\;{\rm if\;the\;constraints\;in\;P1\;hold}\\
	\infty ,{\rm otherwise}
\end{array} \right.$.
Recognizing the constraints in P1 are compact constraint sets, ${\mathbbm{I}_2}\left( {\bm \varphi}  \right)$ is a lower semi-continuous function~\cite{kiwiel:01}, and consequently we have
\begin{equation}\label{eq:33}
	\mathop {\lim \inf }\limits_{j \to \infty } {\mathbbm{I}_2}\left( {{\bm \varphi} _\gamma ^{{n_j}}} \right) \ge {\mathbbm{I}_2}\left( {{\bm \varphi} _\gamma ^ * } \right).
\end{equation}
Taking $j \to \infty $ in \eqref{eq:32} and applying \eqref{eq:33}, we obtain
\begin{equation}\label{eq:34}
	{\rm{{\cal G}}}\left( {{\bm v}_\gamma ^ * ,{\bm \varphi} } \right) + {\mathbbm{I}_2}\left( {\bm \varphi}  \right) \ge {\rm{{\cal G}}}\left( {{\bm v}_\gamma ^ * ,{\bm \varphi} _\gamma ^ * } \right) + {\mathbbm{I}_2}\left( {{\bm \varphi} _\gamma ^ * } \right),\;\;\forall {\bm \varphi},
\end{equation}  
which guarantees that
\begin{equation}\label{eq:35}
	{\bm 0} \in {\nabla _{\bm \varphi} }{\rm{{\cal G}}}\left( {{\bm v}_\gamma ^ * ,{\bm \varphi} _\gamma ^ * } \right) + {\partial _{\bm \varphi} }{\mathbbm{I}_2}\left( {{\bm \varphi} _\gamma ^ * } \right).
\end{equation}

Since ${\mathbbm{I}_1}\left( {{\bm p},{\bm v}} \right)$ does not depend on $\bm \varphi$, we have ${\bm 0} \in {\partial _{\bm \varphi} }{\mathbbm{I}_1}\left( {{\bm p^*_\gamma},{\bm v^*_\gamma}} \right)$. Using similar arguments, we also have ${\bm 0} \in {\nabla _{\bm \varphi} }{\rm{{\cal F}}}\left( {{\bm p}_\gamma ^ * ,{\bm v}_\gamma ^ * } \right)$, ${\bm 0} \in {\nabla _{\bm p}}{\rm{{\cal G}}}\left( {{{\bm v}^ *_\gamma },{{\bm \varphi} ^ *_\gamma }} \right)$, ${\bm 0} \in {\partial _{\bm p}}{\mathbbm{I}_2}\left( {{{\bm \varphi} ^ *_\gamma }} \right)$ and ${\bm 0} \in {\partial _{\bm v}}{\mathbbm{I}_2}\left( {{{\bm \varphi} ^ *_\gamma }} \right)$. Based on these and combine with \eqref{eq:31} and \eqref{eq:35}, we obtain \eqref{eq:36}.
\begin{figure*}
	\begin{equation}\label{eq:36}
		{\bm 0} \in \left[ {\begin{array}{*{20}{c}}
				{{\nabla _{\bm p}}{\rm{{\cal F}}}\left( {{\bm p}_\gamma ^*,{\bm v}_\gamma ^*} \right) + {\nabla _{\bm p}}{\rm{{\cal G}}}\left( {{{\bm v}^ *_\gamma },{{\bm \varphi} ^ *_\gamma }} \right) + {\partial _{\bm p}}{\mathbbm{I}_1}\left( {{\bm p}_\gamma ^*,{\bm v}_\gamma ^*} \right)+{\partial _{\bm p}}{\mathbbm{I}_2}\left( {{{\bm \varphi} ^ *_\gamma }} \right)}\\
				{{\nabla _{\bm v}}{\rm{{\cal F}}}\left( {{\bm p}_\gamma ^*,{\bm v}_\gamma ^*} \right) + {\nabla _{\bm v}}{\rm{{\cal G}}}\left( {{\bm v}_\gamma ^*,{\bm \varphi} _\gamma ^*} \right) + {\partial _{\bm v}}{\mathbbm{I}_1}\left( {{\bm p}_\gamma ^*,{\bm v}_\gamma ^*} \right)+{\partial _{\bm v}}{\mathbbm{I}_2}\left( {{{\bm \varphi} ^ *_\gamma }} \right)}\\
				{{\nabla _{\bm \varphi} }{\rm{{\cal F}}}\left( {{\bm p}_\gamma ^ * ,{\bm v}_\gamma ^ * } \right)+\nabla _{\bm \varphi} }{\rm{{\cal G}}}\left( {{\bm v}_\gamma ^ * ,{\bm \varphi} _\gamma ^ * } \right) + {\partial _{\bm \varphi} }{\mathbbm{I}_1}\left( {{\bm p^*_\gamma},{\bm v^*_\gamma}} \right) + {\partial _{\bm \varphi} }{\mathbbm{I}_2}\left( {{\bm \varphi} _\gamma ^ * } \right)
		\end{array}} \right].
	\end{equation}
	\rule[-10pt]{18.07cm}{0.1em}
\end{figure*}
Therefore, this limit point ${\left\{ {{\bm p}_\gamma ^*,{\bm v}_\gamma ^*,{\bm \varphi} _\gamma ^*} \right\}}$ is a stationary point of \eqref{eq:5}~\cite{bolte:14}. Part 1) of \textbf{Proposition 1} is thus proved.

With $\mathop {\lim }\limits_{j \to \infty } \left( {{\bm p}_\gamma ^{{n_j}},{\bm v}_\gamma ^{{n_j}},{\bm \varphi} _\gamma ^{{n_j}}} \right) = \left( {{\bm p}_\gamma ^ * ,{\bm v}_\gamma ^ * ,{\bm \varphi} _\gamma ^ * } \right)$, and the sequence of solutions ${\left\{ {{\bm p}_\gamma ^{n_j},{\bm v}_\gamma ^{n_j},{\bm \varphi} _\gamma ^{n_j}} \right\}_{j \in {\rm{{\mathbb N}}}}}$ are bounded for any $\gamma$, we know ${\left\{ {{\bm p}_\gamma ^*,{\bm v}_\gamma ^*,{\bm \varphi} _\gamma ^*} \right\}}$ is also bounded for any $\gamma$. Then there must exist a subsequence ${\left\{ {{\bm p}_{\gamma_l} ^*,{\bm v}_{\gamma_l} ^*,{\bm \varphi} _{\gamma_l} ^*} \right\}}_{l \in {\rm{{\mathbb N}}}}$ converges to the limit point $\left\{ {{\bm p}_\infty ^ * ,{\bm v}_\infty ^ * ,{\bm \varphi} _\infty ^ * } \right\}$ based on Bolzano-Weierstrass Theorem~\cite{bartle:00}.

Suppose that ${{\bm v}_\infty ^ *  \ne {\bm \varphi} _\infty ^ * }$, we would have ${{{\left| {{{\bm v}^{\rm{*}}_{\gamma_l}} - {{\bm \varphi} ^{\rm{*}}_{\gamma_l}}} \right|}_2^2}}>c_0$ for a positive $c_0$ and sufficiently large $l$. That yields $\mathop {\lim }\limits_{l \to \infty } {\rm{{\cal G}}}\left( {{\bm v}_{{\gamma _l}}^ * ,{\bm \varphi} _{{\gamma _l}}^ * } \right) = \mathop {\lim }\limits_{l \to \infty } \frac{{{\gamma _l}}}{2}{{\left| {{\bm v}_{{\gamma _l}}^{ * } - {\bm \varphi} _{{\gamma _l}}^{ * }} \right|}_2^2} = \infty $, which violates the bounded property obtained below \eqref{eq:28}. Hence, by contradiction, we must have ${{\bm v}_\infty ^ *  = {\bm \varphi} _\infty ^ * }$. According to part 1) of the \textbf{Proposition 1}, $\left\{ {{\bm p}_\gamma ^ * ,{\bm v}_\gamma ^ * ,{\bm \varphi} _\gamma ^ * } \right\}$ is a stationary point of \eqref{eq:5} for any $\gamma$. Thus we have $\left\{ {{\bm p}_\infty ^ * ,{\bm v}_\infty ^ * ,{\bm \varphi} _\infty ^ * } \right\}$ is a stationary point of \eqref{eq:5}. Together with ${{\bm v}_\infty ^ *  = {\bm \varphi} _\infty ^ * }$, which means that \eqref{eq:5} is equivalent to \eqref{eq:4}, we obtain that $\left\{ {{\bm p}_\infty ^ * ,{\bm v}_\infty ^ *} \right\}$ is a stationary point of \eqref{eq:4}. This completed the proof of part 2).

\subsection{Proof of Lemma 1}
Firstly, we discuss the ES and TS STAR-RIS case. Since ${\varphi _m^{\mathcal t}}$ and ${\varphi _m^{\mathcal r}}$ are separable in both the objective function and constraints, we can consider one ${\varphi _m^{\mathcal t}}$ or ${\varphi _m^{\mathcal r}}$ at a time. Taking ${\varphi _m^{\mathcal t}}$ as an example, the optimization problem \eqref{eq:7} with respect to ${\varphi _m^{\mathcal t}}$ is 
\begin{subequations}\label{eq:37}
	\begin{align}
		\mathop {\min }\limits_{\varphi _m^{\mathcal t}}\;\;& {\left| {\varphi _m^{\mathcal t}} \right|^2} - 2\left| {v_m^{\mathcal t}} \right|\left| {\varphi _m^{\mathcal t}} \right|\cos \left( {\angle \varphi _m^{\mathcal t} - \angle v_m^{\mathcal t}} \right)\label{eq:37a}\\
		s.t.\;\;&\left| {\varphi _m^{\mathcal t}} \right| = 1,\;\;\;\;{\rm if\;TS}\label{eq:37b}\\
		&\angle \varphi _m^{\mathcal t} \in \left\{ {0,{{2\pi } \mathord{\left/
					{\vphantom {{2\pi } L}} \right.
					\kern-\nulldelimiterspace} L}, \cdots ,{{2\pi \left( {L - 1} \right)} \mathord{\left/
					{\vphantom {{2\pi \left( {L - 1} \right)} L}} \right.
					\kern-\nulldelimiterspace} L}} \right\},\label{eq:37c}
	\end{align}
\end{subequations} 
where we have expanded ${\left| {v_m^{\mathcal t} - \varphi _m^{\mathcal t}} \right|^2}$ and removed terms not related to ${\varphi _m^{\mathcal t}}$. To minimize \eqref{eq:37a}, $\cos \left( {\angle \varphi _m^{\mathcal t} - \angle v_m^{\mathcal t}} \right)$ should be maximized. Taking the consideration of \eqref{eq:37c}, the optimal phase of $\angle \varphi _m^{{\mathcal t}}$ is equal to ${{\rm Proj}_{{\bm \Theta }}}\left( {\angle v_m^{\mathcal t}} \right)$. Since TS mode restrict the amplitude of $\varphi_m^{\mathcal t}$, ${{\rm Proj}_{{\bm \Theta }}}\left( {\angle v_m^{\mathcal t}} \right)$ is the solution of the TS mode. For the ES STAR-RIS, we put this optimal phase back into the objective function \eqref{eq:37a}, the resulting problem with respect to $\left| {\varphi _m^{\mathcal t}} \right|$ is an unconstrained quadratic optimization problem and the closed-form solution is shown in \textbf{Lemma 1}.

For the MS STAR-RIS, there are two possibilities for the amplitude variables.

1) $\left| {\varphi _m^{\mathcal t}} \right| = 0$ and $\left| {\varphi _m^{\mathcal r}} \right| = 1$. Through expanding the objective function in \eqref{eq:7a} and removing the terms unrelated to ${\varphi _m^{\mathcal r}}$, the following optimization problem can be obtained from \eqref{eq:7}:
\begin{equation}\label{eq:38}
	\begin{split}
		\mathop {\min }\limits_{\angle \varphi _m^{\mathcal r}}&\;\; - 2\left| {v_m^{\mathcal r}} \right|\cos \left( {\angle \varphi _m^{\mathcal r} - \angle v_m^{\mathcal r}} \right)\\
		s.t.&\;\;\angle \varphi _m^{\mathcal r} \in \left\{ {0,{{2\pi } \mathord{\left/
					{\vphantom {{2\pi } L}} \right.
					\kern-\nulldelimiterspace} L}, \cdots ,{{2\pi \left( {L - 1} \right)} \mathord{\left/
					{\vphantom {{2\pi \left( {L - 1} \right)} L}} \right.
					\kern-\nulldelimiterspace} L}} \right\}.
	\end{split}
\end{equation}
Therefore, the optimal solution of $\angle \varphi _m^{{\mathcal r}}$ is ${{\rm Proj}_{{\bm \Theta }}}\left( {\angle v_m^{\mathcal r}} \right)$, and the minimal value is $- 2\left| {v_m^{\mathcal r}} \right|$ $\cos \left( {{{\rm Proj}_{{\bm \Theta }}}\left( {\angle v_m^{\mathcal r}} \right) - \angle v_m^{\mathcal r}} \right)=- 2\beta _m^{\mathcal r}$.

2) $\left| {\varphi _m^{\mathcal t}} \right| = 1$ and $\left| {\varphi _m^{\mathcal r}} \right| = 0$. With similar derivations to the above case, the optimal solution of $\angle \varphi _m^{{\mathcal t}}$ is ${{\rm Proj}_{{\bm \Theta }}}\left( {\angle v_m^{\mathcal t}} \right)$. Hence, the minimal value is $- 2\left| {v_m^{\mathcal t}} \right|\cos \left( {{{\rm Proj}_{{\bm \Theta }}}\left( {\angle v_m^{\mathcal t}} \right) - \angle v_m^{\mathcal t}} \right)=- 2\beta _m^{\mathcal t}$.

Finally, the optimal solution can be obtained by choosing the minimal value of the above two cases and the result is expressed in \textbf{Lemma 1} using ${\mathop{\rm sgn}} \left(  \cdot  \right)$ function.

\subsection{Proof of Lemma 2}
Let $\varphi _m^{\mathcal t} = a_m^{\mathcal t}{e^{j\theta _m^{\mathcal t}}}$, where $a_m^{\mathcal t} \in {\mathbb{R}}$ and $\theta _m^{\mathcal t} \in \left\{ {0,{{2\pi } \mathord{\left/
			{\vphantom {{2\pi } L}} \right.
			\kern-\nulldelimiterspace} L}, \cdots ,{{2\pi \left( {L - 1} \right)} \mathord{\left/
			{\vphantom {{2\pi \left( {L - 1} \right)} L}} \right.
			\kern-\nulldelimiterspace} L}} \right\}$. By expanding \eqref{eq:9a} and removing the terms irrelvant to $\varphi _m^{\mathcal t}$ and $\varphi _m^{\mathcal r}$, we have the optimization problem \eqref{eq:39}.
\begin{figure*}
	\begin{equation}\label{eq:39}
		\begin{split}
			\mathop {\min }\limits_{a_m^{\mathcal t},\theta _m^{\mathcal t},\varphi _m^{\mathcal r}} \;\;&{\left( {a_m^{\mathcal t}} \right)^2} - 2a_m^{\mathcal t}\left| {v_m^{\mathcal t}} \right|\cos \left( {\theta _m^{\mathcal t} - \angle v_m^{\mathcal t}} \right) + {\left| {\varphi _m^{\mathcal r}} \right|^2} - 2\left| {v_m^{\mathcal r}} \right|\left| {\varphi _m^{\mathcal r}} \right|{\mathop{\rm Re}\nolimits} \left\{ {{e^{j\left( {\angle \varphi _m^{\mathcal r} - \angle v_m^{\mathcal r}} \right)}}} \right\}\\
			s.t.\;\;&\angle \varphi _m^{\mathcal r} = \theta _m^{\mathcal t} \pm {\pi  \mathord{\left/
					{\vphantom {\pi  2}} \right.
					\kern-\nulldelimiterspace} 2}\left( {\bmod 2\pi } \right),\\
			\;\;&\theta _m^{\mathcal t} \in \left\{ {0,{{2\pi } \mathord{\left/
						{\vphantom {{2\pi } \L}} \right.
						\kern-\nulldelimiterspace} L}, \cdots ,{{2\pi \left( {L - 1} \right)} \mathord{\left/
						{\vphantom {{2\pi \left( {L - 1} \right)} L}} \right.
						\kern-\nulldelimiterspace} L}} \right\}.
		\end{split}
	\end{equation}
\end{figure*}

To minimize \eqref{eq:39}, $\angle \varphi _m^{\mathcal r}$ should be chosen to maximize the term ${\mathop{\rm Re}\nolimits} \left\{ {{e^{j\left( {\angle \varphi _m^{\mathcal r} - \angle v_m^{\mathcal r}} \right)}}} \right\}$. On the other hand, from the constraint in \eqref{eq:39}, once ${\theta _m^{\mathcal t}}$ is obtained, there are only two possible values for ${\angle \varphi _m^{\mathcal r}}$ to choose from. Hence we have
\begin{equation}\label{eq:40}
	\angle \varphi _m^{\mathcal r} = \left\{ \begin{array}{l}
		\theta _m^{\mathcal t} + {\pi  \mathord{\left/
				{\vphantom {\pi  2}} \right.
				\kern-\nulldelimiterspace} 2},\;\;{\rm if}\;\;\sin \left( {\theta _m^{\mathcal t} - \angle v_m^{\mathcal r}} \right) \le 0,\\
		\theta _m^{\mathcal t} - {\pi  \mathord{\left/
				{\vphantom {\pi  2}} \right.
				\kern-\nulldelimiterspace} 2},\;\;{\rm otherwise}.
	\end{array} \right.
\end{equation}
Putting \eqref{eq:40} into \eqref{eq:39}, the problem is reduced to \eqref{eq:41}.
\begin{figure*}
	\begin{equation}\label{eq:41}
		\begin{split}
			\mathop {\min }\limits_{a_m^{\mathcal t},\theta _m^{\mathcal t},\left| {\varphi _m^{\mathcal r}} \right|}& {\left( {a_m^{\mathcal t}} \right)^2} - 2a_m^{\mathcal t}\left| {v_m^{\mathcal t}} \right|\cos \left( {\theta _m^{\mathcal t} - \angle v_m^{\mathcal t}} \right) + {\left| {\varphi _m^{\mathcal r}} \right|^2} - 2\left| {v_m^{\mathcal r}} \right|\left| {\varphi _m^{\mathcal r}} \right|\left| {\sin \left( {\theta _m^{\mathcal t} - \angle v_m^{\mathcal r}} \right)} \right|\\
			s.t.\;\;&\theta _m^{\mathcal t} \in \left\{ {0,{{2\pi } \mathord{\left/
						{\vphantom {{2\pi } L}} \right.
						\kern-\nulldelimiterspace} L}, \cdots ,{{2\pi \left( {L - 1} \right)} \mathord{\left/
						{\vphantom {{2\pi \left( {L - 1} \right)} L}} \right.
						\kern-\nulldelimiterspace} L}} \right\} .
		\end{split}
	\end{equation}
	\rule[-10pt]{18.07cm}{0.1em}
\end{figure*}
This is an unconstrained quadratic function for ${a_m^{\mathcal t}}$ and ${\left| {\varphi _m^{\mathcal r}} \right|}$, hence their optimal solutions are
\begin{equation}\label{eq:42}
	\left\{ \begin{array}{l}
		a_m^{\mathcal t} = \left| {v_m^{\mathcal t}} \right|\cos \left( {\theta _m^{\mathcal t} - \angle v_m^{\mathcal t}} \right),\\
		\left| {\varphi _m^{\mathcal r}} \right| = \left| {v_m^{\mathcal r}} \right|\left| {\sin \left( {\theta _m^{\mathcal t} - \angle v_m^{\mathcal r}} \right)} \right|.
	\end{array} \right.
\end{equation}
With the second line of \eqref{eq:42} and using \eqref{eq:40}, we obtain the solution of $\varphi _m^{\mathcal r}$.

Finally, substituting \eqref{eq:42} back to \eqref{eq:41}, the resulting problem with respect to $\theta _m^{\mathcal t}$ is
\begin{equation}\label{eq:43}
	\begin{split}
		\mathop {\min }\limits_{\theta _m^{\mathcal t}}\;\;&  - {\left| {v_m^{\mathcal t}} \right|^2}{\cos ^2}\left( {\theta _m^{\mathcal t} - \angle v_m^{\mathcal t}} \right) - {\left| {v_m^{\mathcal r}} \right|^2}{\sin ^2}\left( {\theta _m^{\mathcal t} - \angle v_m^{\mathcal r}} \right).\\
		s.t.\;\;&\theta _m^{\mathcal t} \in \left\{ {0,{{2\pi } \mathord{\left/
					{\vphantom {{2\pi } L}} \right.
					\kern-\nulldelimiterspace} L}, \cdots ,{{2\pi \left( {L - 1} \right)} \mathord{\left/
					{\vphantom {{2\pi \left( {L - 1} \right)} L}} \right.
					\kern-\nulldelimiterspace} L}} \right\}.
	\end{split}
\end{equation}
Since ${\cos ^2}u = \left( {1 + \cos 2u} \right)/2$ and ${\sin ^2}u = \left( {1 - \cos 2u} \right)/2$, optimizing \eqref{eq:43} is equivalent to 
\begin{equation}\label{eq:44}
	\begin{split}
		\mathop {\min }\limits_{\theta _m^{\mathcal t}}\;\;& {\left| {v_m^{\mathcal r}} \right|^2}\cos \left( {2\theta _m^{\mathcal t} - 2\angle v_m^{\mathcal r}} \right) - {\left| {v_m^{\mathcal t}} \right|^2}\cos \left( {2\theta _m^{\mathcal t} - 2\angle v_m^{\mathcal t}} \right)\\
		s.t.\;\;&\theta _m^{\mathcal t} \in \left\{ {0,{{2\pi } \mathord{\left/
					{\vphantom {{2\pi } L}} \right.
					\kern-\nulldelimiterspace} L}, \cdots ,{{2\pi \left( {L - 1} \right)} \mathord{\left/
					{\vphantom {{2\pi \left( {L - 1} \right)} L}} \right.
					\kern-\nulldelimiterspace} L}} \right\}.
	\end{split}
\end{equation}
Using sum-difference-product formula for trigonometric functions, the objective function in \eqref{eq:44} is equal to
\begin{equation}\label{eq:45}
	{{\sqrt {\chi _m} }}\cos \left( {2\theta _m^{\mathcal t} - 2\angle v_m^{\mathcal t}+{b_m} } \right),
\end{equation}
where ${\chi _m} = {{\left[ {{{\left| {v_m^{\mathcal r}} \right|}^2}\cos \left( {2\angle v_m^{\mathcal t} - 2\angle v_m^{\mathcal r}} \right) + {{\left| {v_m^{\mathcal t}} \right|}^2}} \right]}^2} + {{\left[ {{{\left| {v_m^{\mathcal r}} \right|}^2}\sin \left( {2\angle v_m^{\mathcal t} - 2\angle v_m^{\mathcal r}} \right)} \right]}^2}$ and $b_m$ is defined in \textbf{Lemma 2}. Since ${\chi _m}$ is unrelated to $\theta _m^{\mathcal t}$, \eqref{eq:45} is equivalent to
\begin{equation}\label{eq:46}
	\begin{split}
		\mathop {\min }\limits_{\theta _m^{\mathcal t}}\;\;&\cos \left( {2\theta _m^{\mathcal t} - 2\angle v_m^{\mathcal t}+{b_m} } \right)\\
		s.t.\;\;&\theta _m^{\mathcal t} \in \left\{ {0,{{2\pi } \mathord{\left/
					{\vphantom {{2\pi } L}} \right.
					\kern-\nulldelimiterspace} L}, \cdots ,{{2\pi \left( {L - 1} \right)} \mathord{\left/
					{\vphantom {{2\pi \left( {L - 1} \right)} L}} \right.
					\kern-\nulldelimiterspace} L}} \right\}.
	\end{split}
\end{equation}
Hence, the optimal solution of $\theta _m^{{\mathcal t}}$ is ${{\rm Proj}_{{\bm \Theta }}}\left( {\angle v_m^{\mathcal t} {{ - {b_m}} \mathord{\left/
			{\vphantom {{ - {b_m}} 2}} \right.
			\kern-\nulldelimiterspace} 2} + {\pi  \mathord{\left/
			{\vphantom {\pi  2}} \right.
			\kern-\nulldelimiterspace} 2}} \right)$ and the optimal solution of \eqref{eq:9} shown in \textbf{Lemma 2}.

\subsection{Proof of Lemma 3}
Firstly, to deal with the sum-of-logarithms-of-ratio objective function in \eqref{eq:12}, the closed-form fractional programming (FP) approach in~\cite{Shen:18,Guo:20} is introduced, which has two key steps.

1) \emph{Lagrangian Dual Transform:} The logarithm function can be represented with an auxiliary variable $\rho$ as
\begin{equation}\label{eq:47}
	\log \left( {1 + \gamma } \right) = \mathop {\max }\limits_{\rho} \log \left( {1 + \rho } \right) - \rho  + \frac{{\left( {1 + \rho } \right)  \gamma }}{{1 + \gamma }},
\end{equation}
where the optimal solution occurs at $\rho = \gamma$. Based on \eqref{eq:47}, the objective function \eqref{eq:12} can be written as
\begin{equation}\label{eq:48}
	\begin{split}
		&{\cal R}\left( {{\bm w},{{\bm v}^{\mathcal t}},{{\bm v}^{\mathcal r}},{\lambda^{\mathcal t}},{\lambda^{\mathcal r}}} \right) =\\
		&\mathop {\max }\limits_{\rho}\;{\lambda^{\mathcal r}}\sum\limits_{l = 1}^{{K^{\mathcal r}}} {\log \left( {1 + {\rho _l}} \right) - {\rho _l} + \frac{{\left( {1 + {\rho _l}} \right){{\left| {{\bm a}_l^T{{\bm w}_l}} \right|}^2}}}{{\sum\nolimits_{i = 1}^{{K^{\mathcal r}} + {K^{\mathcal t}}} {{{\left| {{\bm a}_l^T{{\bm w}_i}} \right|}^2}}  + {\lambda^{\mathcal r}}\sigma _l^2}}} \\
		&+ {\lambda^{\mathcal t}}\sum\limits_{l = {K^{\mathcal r}} + 1}^{{K^{\mathcal r}} + {K^{\mathcal t}}} {\log \left( {1 + {\rho _l}} \right) - {\rho _l} + \frac{{\left( {1 + {\rho _l}} \right){{\left| {{\bm a}_l^T{{\bm w}_l}} \right|}^2}}}{{\sum\nolimits_{i = 1}^{{K^{\mathcal r}} + {K^{\mathcal t}}} {{{\left| {{\bm a}_l^T{{\bm w}_i}} \right|}^2}}  + {\lambda^{\mathcal t}}\sigma _l^2}}}. 
	\end{split}
\end{equation}

The second step is to tackle the fractional term in \eqref{eq:48}.

2) \emph{Quadratic transform:} By introducing the auxiliary variable $x$, the following equation holds.
\begin{equation}\label{eq:49}
	\frac{{{{\left| {A\left( {\bm u} \right)} \right|}^2}}}{{B\left( {\bm u} \right)}} = 2{\mathop{\rm Re}\nolimits} \left\{ {\overline x A\left( {\bm u} \right)} \right\} - {\left| x \right|^2}B\left( {\bm u} \right).
\end{equation}
The equivalence can be proved by substituting $x = {{A\left( {\bm u} \right)} \mathord{\left/
		{\vphantom {{A\left( {\bm u} \right)} {B\left( {\bm u} \right)}}} \right.
		\kern-\nulldelimiterspace} {B\left( {\bm u} \right)}}$. With the transformation in \eqref{eq:49}, \eqref{eq:48} can be converted into
\begin{equation}\label{eq:50}
	{\rm{{\cal R}}}\left( {{\bm w},{{\bm v} ^{\mathcal t}},{{\bm v} ^{\mathcal r}},{{\lambda} ^{\mathcal t}},{{\lambda} ^{\mathcal r}}} \right) = \mathop {\max }\limits_{{\bm \rho},{\bm x}} {{\rm{{\cal F}}}_1}\left( {{\bm w},{\bm \rho},{\bm x},{{\bm v} ^{\mathcal t}},{{\bm v} ^{\mathcal r}},{{\lambda} ^{\mathcal t}},{{\lambda} ^{\mathcal r}} } \right),
\end{equation}
where ${{\rm{{\cal F}}}_1}$ is defined in \eqref{eq:13}.
\subsection{Proof of Lemma 4}
Introducing the dual variable $\mu $ to the constraint \eqref{eq:17b}, the problem \eqref{eq:17} is equivalent to
\begin{equation}\label{eq:51}
	\mathop {\max }\limits_{{\mu } \ge 0} \mathop {\min }\limits_{\bm w} \sum\limits_{l = 1}^{K^{\mathcal r} + K^{\mathcal t}} { {{\bm w}_l^H{{\bm \Xi}}{{\bm w}_l} - 2{\mathop{\rm Re}\nolimits} \left[ {{\bm q}_l^H  {{\bm w}_l}} \right] + {\mu }\left( {{\bm w}_l^H{{\bm w}_l} - {P_{BS}}} \right)} }.
\end{equation}
For each fixed $\mu $, this problem is a quadratic function for all $\left\{ {{{\bm w}_l}} \right\}_{l = 1}^{K^{\mathcal r} + K^{\mathcal t}}$. Therefore, the optimal solution of ${{\bm w}_l}$ is ${\left( {{\bm \Xi} + \mu {\bm I}} \right)^{ - 1}}{{\bm q}_l}$. Through eigenvalue decomposition ${\bm \Xi} = {\bm U}{\bm \Lambda} {{\bm U}^H}$, the optimal ${{\bm w}_l}$ can be rewritten as 
\begin{equation}\label{eq:52}
	{{\bm w}_l} = {\bm U}{\left( {{\bm \Lambda}  + \mu {\bm I}} \right)^{ - 1}}{{\bm U}^H}{{\bm q}_l}.
\end{equation}
According to the complementary slackness, there are two cases for $\mu$:

1) $\mu = 0$. That means the constraint \eqref{eq:17b} is inactive. This happens when $\Sigma_{l = 1}^{K^{\mathcal r} + K^{\mathcal t}} {{\bm w}_l^H{{\bm w}_l} \le {P_{BS}}} $ with ${{\bm w}_l} = {\bm U}{{\bm \Lambda} ^{ - 1}}{{\bm U}^H}{{\bm q}_l}$, which is equivalent to $\Sigma_{l = 1}^{K^{\mathcal r} + K^{\mathcal t}} {{\bm q}_l^H{\bm U}{{\bm \Lambda} ^{ - 2}}{{\bm U}^H}{{\bm q}_l}} \le {P_{BS}}$. Recall the definition of $\bm B$ in \textbf{Lemma 4}, $\Sigma_{l = 1}^{K^{\mathcal r} + K^{\mathcal t}} {{\bm q}_l^H{\bm U}{{\bm \Lambda} ^{ - 2}}{{\bm U}^H}{{\bm q}_l}} = {\rm Tr}\left( {{{\bm \Lambda} ^{ - 2}}{\bm B}} \right)$. Hence, this case only occurs when ${\rm Tr}\left( {{{\bm \Lambda} ^{ - 2}}{\bm B}} \right) \le {P_{BS}}$.

2) $\mu > 0$. That means the constraint \eqref{eq:17b} is active. Therefore, we need to solve the equation 
$\Sigma_{l = 1}^{K^{\mathcal r} + K^{\mathcal t}} {{\bm w}_l^H{{\bm w}_l} = {P_{BS}}} $ with each ${\bm w}_l$ given by \eqref{eq:52}. Putting \eqref{eq:52} into the constraint $\Sigma_{l = 1}^{K^{\mathcal r} + K^{\mathcal t}} {{\bm w}_l^H{{\bm w}_l} = {P_{BS}}} $ and rearrange the terms with the trace operator, it can be shown that 
\begin{equation}\label{eq:53}
	{\rm Tr}\left( {{{\left( {{\bm \Lambda}  + \mu {\bm I}} \right)}^{ - 2}}{\bm B}} \right) = {P_{BS}}.
\end{equation}

Notice that the left hand side expression is a monotonic decreasing function of $\mu$. The bisection method with searching interval $\left[ {0,{\mu _{\max }}} \right]$ can be adopted to find this solution, where ${\mu _{\max }}$ should satisfies
\begin{equation}\label{eq:54}
	{\rm Tr}\left( {{{\left( {{\bm \Lambda}  + {\mu _{\max }} {\bm I}} \right)}^{ - 2}}{\bm B}} \right) \le {P_{BS}}.
\end{equation}
Since $\bm \Xi$ is a positive semidefinite matrix, the diagonal element of $\bm \Lambda$ is non-negative. Hence, if ${\rm Tr}\left( {{{\left( {{\mu _{\max }} {\bm I}} \right)}^{ - 2}}{\bm B}} \right) \le {P_{BS}}$ holds, we would also have \eqref{eq:54} holds. From ${\rm Tr}\left( {{{\left( {{\mu _{\max }} {\bm I}} \right)}^{ - 2}}{\bm B}} \right) \le {P_{BS}}$, we can set ${\mu _{\max }} = \sqrt {{{{\rm Tr}\left( {\bm B} \right)} \mathord{\left/
			{\vphantom {{{\rm Tr}\left( {\bm B} \right)} {{P_{BS}}}}} \right.
			\kern-\nulldelimiterspace} {{P_{BS}}}}} $.

\end{document}